\pdfoutput=1

\documentclass[11pt,twoside,a4paper,cmspaper,final,collab]{cms-tdr}
\def\svnVersion{365277}\def\cmsCernNoTag{CERN-EP/2016-043}\def\cmsCernDate{\today}\def\cmsMessage{Published in the European Physical Journal C as \href{http://dx.doi.org/10.1140/epjc/s10052-016-4293-4}{\doi{10.1140/epjc/s10052-016-4293-4.}}}

\begin{document}\cmsNoteHeader{SMP-14-022}

\hyphenation{had-ron-i-za-tion}
\hyphenation{cal-or-i-me-ter}
\hyphenation{de-vices}
\RCS$Revision: 354363 $
\RCS$HeadURL: svn+ssh://svn.cern.ch/reps/tdr2/papers/SMP-14-022/trunk/SMP-14-022.tex $
\RCS$Id: SMP-14-022.tex 354363 2016-07-10 16:42:34Z alverson $
\newlength\cmsFigWidth
\ifthenelse{\boolean{cms@external}}{\setlength\cmsFigWidth{0.98\columnwidth}}{\setlength\cmsFigWidth{0.45\textwidth}}
\newlength\cmsFigWidthNarrow
\ifthenelse{\boolean{cms@external}}{\setlength\cmsFigWidthNarrow{0.80\columnwidth}}{\setlength\cmsFigWidthNarrow{0.45\textwidth}}
\ifthenelse{\boolean{cms@external}}{\providecommand{\cmsLeft}{top\xspace}}{\providecommand{\cmsLeft}{left\xspace}}
\ifthenelse{\boolean{cms@external}}{\providecommand{\cmsRight}{bottom\xspace}}{\providecommand{\cmsRight}{right\xspace}}
\providecommand{\CL}{CL\xspace}

\newcommand{\sigeta}{\ensuremath{\sigma_\eta}\xspace}
\newcommand{\difsig}{\ensuremath{\frac{ \rd\sigma}{\rd\eta}}\xspace}
\newcommand{\zmumu}{\ensuremath{\cPZ/\gamma^* \to \Pgmp\Pgmm}\xspace}
\newcommand{\ztautau}{\ensuremath{\cPZ/\gamma^*\to\PGt^+ \PGt^-}\xspace}
\newcommand{\RQCD}{\ensuremath{R_{\pm}^{\mathrm{QCD}}}}
\newcommand{\etaabs} {\ensuremath{\abs{ \eta }}\xspace}
\hyphenation{Bjor-ken}

\cmsNoteHeader{SMP-14-022}
\title{Measurement of the differential cross section and charge asymmetry for inclusive $\Pp\Pp \to \PW^{\pm}+X$ production at $\sqrt{s} = 8\TeV$}
\titlerunning{Differential cross section and charge asymmetry for inclusive $\Pp\Pp \to \PW^{\pm}+X$ production at 8\TeV}

\date{\today}

\abstract{
The differential cross section and charge asymmetry for inclusive
$\Pp\Pp \to \PW^{\pm}+X \to \PGm^{\pm}\PGn+X$
production at $\sqrt{s}=8\TeV$ are measured as a function of muon pseudorapidity.
The data sample corresponds to an integrated luminosity of 18.8\fbinv recorded with the CMS detector at the LHC.
These results provide important constraints on the parton distribution functions of the proton in the range of the Bjorken scaling variable $x$ from $10^{-3}$ to $10^{-1}$.
}

\hypersetup{%
pdfauthor={CMS Collaboration},%
pdftitle={Measurement of the differential cross sections and charge asymmetry of inclusive pp to WX production at sqrt(s)=8 TeV},%
pdfsubject={CMS},%
pdfkeywords={CMS, physics, W boson, muon charge asymmetry}}

\maketitle

\section{Introduction}
We present measurements of the $\Pp\Pp \to \Wpm+X \to \PGm^{\pm}\PGn+X$ differential cross section and the muon charge asymmetry that provide important constraints on the valence and sea quark distributions in the proton. Uncertainties in the parton distribution functions (PDF) have become a limiting factor for the precision of many inclusive and differential cross section calculations, given the development of precise theoretical tools describing hard scattering processes in $\Pp\Pp$ collisions.

{\tolerance=5000
For each charge of the $\PW$ boson, the differential cross section,
\begin{equation}
    \sigma^{\pm}_{\eta}=\difsig(\Pp\Pp \to \Wpm+X \to \PGm^{\pm}\PGn+X),
\end{equation}
is measured in bins of muon pseudorapidity $\eta = -\ln \tan(\theta/2)$ in the laboratory frame, where $\theta$ is the polar angle of the muon direction with respect to the beam axis. Current theoretical calculations predict these cross sections with next-to-next-to-leading-order (NNLO) accuracy in perturbative quantum chromodynamics (QCD). The dominant $\Wpm$ boson production occurs through the annihilation of a valence quark from one of the protons with a sea antiquark from the other: $\cPqu\cPaqd \to \PWp$ and $\cPqd\cPaqu \to \PWm$. Because of the presence of two valence $\cPqu$ quarks in the proton, $\PWp$ bosons are produced more often than $\PWm$ bosons. Precise measurement of the charge asymmetry as a function of the muon $\eta$,
\begin{equation}
    \mathcal{A}(\eta) = \frac{\sigma^+_{\eta}-\sigma^-_{\eta}}{\sigma^+_{\eta}+\sigma^-_{\eta}},
\end{equation}
provides significant constraints on the ratio of $\cPqu$ and $\cPqd$ quark distributions in the proton for values of $x$, the Bjorken scaling variable~\cite{Bjorken}, between $10^{-3}$ and $10^{-1}$.
}

The $\Wpm$ boson production asymmetry was previously studied in $\Pp\Pap$ collisions by the CDF and D0 collaborations~\cite{CDF:wasym1, CDF:wasym2, Abazov:2007pm, Abazov:2008qv, Abazov:2013rja}. At the LHC, the first measurements of the lepton charge asymmetries were performed  by the CMS, ATLAS, and LHCb experiments using data collected in 2010~\cite{CMS:asym:2010, PhysRevD.85.072004,  LHCb:wz:2010}. The CMS experiment has further improved the measurement precision in both the electron and muon decay channels using data from $\Pp\Pp$ collisions at $\sqrt{s}=7\TeV$ corresponding to integrated luminosities of 0.84\fbinv and 4.7\fbinv in the electron~\cite{CMS:asym:2011} and muon~\cite{CMS:asym:2012} decay channels, respectively.

This measurement is based on a data sample of $\Pp\Pp$ collisions at $\sqrt{s}=8\TeV$ collected by CMS during 2012,
corresponding to an integrated luminosity of 18.8\fbinv. At $\sqrt{s}=8\TeV$ the average value of the Bjorken scaling variable for
the interacting partons in $\Wpm$ boson production is lower than at $\sqrt{s}=7\TeV$, which is expected to result in a lower $\Wpm$ boson production charge asymmetry.
This measurement provides important constraints on the proton PDFs, which is illustrated by the QCD analysis also presented in this paper.

\section{CMS detector}

The central feature of the CMS apparatus is a superconducting solenoid of 6\unit{m} internal diameter, providing a magnetic field of 3.8\unit{T}.
Within the solenoid volume are a silicon pixel and strip tracker, a lead tungstate crystal electromagnetic calorimeter, and a brass and scintillator hadron calorimeter, each composed of a barrel and two endcap sections. Muons are measured in gas-ionization detectors embedded in the steel flux-return yoke outside the solenoid. Extensive forward calorimetry complements the coverage provided by the barrel and endcap detectors. A more detailed description of the CMS detector can be found in Ref.~\cite{Chatrchyan:2008zzk}.

\section{Data selection and simulation}

A $\Wpm\to\PGm^{\pm}\PGn$ event is characterized by an isolated muon with a high transverse momentum $\PT$ and a large missing transverse energy \ETslash associated with an undetected neutrino. Events in this sample are collected with an isolated single-muon trigger with a $\pt$ threshold of 24\GeV. To reduce the background, identification and isolation criteria are applied to the reconstructed muons. These requirements are similar to those used in the previous measurement~\cite{CMS:asym:2012}. Muon tracks must be reconstructed in both the silicon tracker and the muon detectors. The global muon fit is required to have a $\chi^2$ per degree of freedom less than 10. The pseudorapidity coverage for reconstructed muons is restricted to $\abs{\eta}<2.4$. Cosmic ray contamination is largely reduced by rejecting the muon candidates with a large $({>}0.2\cm)$ distance of closest approach to the primary vertex in the transverse plane. The isolation criterion is based on additional tracks reconstructed in a cone of $\sqrt{\smash[b]{(\Delta\eta)^2+(\Delta\phi)^2}}<0.3$ around the muon, where $\phi$ is the azimuthal angle (in radians) in the laboratory frame. The muon candidate is rejected if the scalar $\pt$ sum of these tracks is more than 10\% of the muon $\pt$. The selected muon candidate with the largest $\pt$, identified as a signal muon from the $\PW$ boson decay, is required to have a $\PT>25\GeV$ and also to be the particle that triggered the event. To reduce the background from Drell--Yan (DY) dimuon production, events containing a second identified muon with $\PT>15\GeV$ are rejected.

A total of about 61 million $\PWp \to \Pgmp \Pgn$ and 45 million $\PWm \to \Pgmm\Pagn$ candidate events are selected. The $\Wpm\to\PGm^{\pm}\PGn$ signal is contaminated with backgrounds that also produce a muon with high $\PT$. The major background sources are (i)~multijet (QCD) events with high-\PT muons produced in hadron decays (about $10\%$ of the selected sample), and (ii)~$\zmumu$ events (5\% of the sample). The contribution from other backgrounds, such as $\Wpm\to\tau^{\pm}\PGn$\,(2.6\%), $\ztautau$\,(0.5\%), and  $\ttbar$\,(0.5\%) events, is relatively small. The contributions from single top quark\,(0.14\%) and diboson\,(0.07\%) events, as well as from cosmic muons\,($10^{-5}$), are negligible.

{\tolerance=5000
Simulated samples are used to model the signal and background processes.
The signal, as well as the electroweak and \ttbar background samples, is based on the next-to-leading-order (NLO) matrix element calculations implemented in the \POWHEG Monte Carlo (MC) event generator~\cite{POWHEG0,POWHEG1,POWHEG2,POWHEG3}, interfaced with \textsc{pythia6}~\cite{PYTHIA6} for parton showering and hadronization, including electromagnetic final-state radiation (FSR).
The CT10 NLO PDFs~\cite{CTEQ:1007} are used.
The $\tau$ lepton decays in relevant processes are simulated with \TAUOLA~\cite{TAUOLA}.
The QCD background is generated with \PYTHIA6 using CTEQ6L PDF~\cite{CTEQ6L}.

The MC events are overlaid by simulated minimum-bias events to model additional $\Pp\Pp$ interactions (pileup) present in data.
The detector response to all generated particles is simulated with \GEANTfour~\cite{GEANT4}. Final-state particles are reconstructed with the same algorithms used for the data sample.
}

\section{Corrections to the data and simulations}

The fiducial cross sections are measured for muon $\pt>25\GeV$ in 11 bins of absolute pseudorapidity, covering the range $\abs{\eta}<2.4$. The $\etaabs$ binning is such that the migration effects due to the finite $\eta$ resolution are negligible. In each $\abs{\eta}$ bin, the number of $\PWp \to \PGm^+\PGn$ and $\PWm \to \PGm^-\PGn$ events is extracted by fitting the $\ETslash$ distributions with signal and background distributions (templates). The template shapes and initial normalizations are derived from MC simulations. To improve the simulation, several corrections are applied to the MC samples. The corrections, which are similar to those used in the previous measurement~\cite{CMS:asym:2012}, are briefly summarized below.

All simulated events are weighted to match the pileup distribution in data. The weight factors are based on the measured instantaneous luminosity and minimum-bias cross section leading to a good description of the average number of reconstructed vertices in the data.

Accurate calibration of the muon momentum is important for the proper modeling of the yields of $\Wpm$ events and of the shapes of $\ETslash$ templates. Dominant sources of the muon momentum mismeasurement are the mismodeling of the tracker alignment and the magnetic field. Muon momentum correction factors are derived using $\zmumu$ events in several iterations~\cite{bib:momcor}. First, ``reference'' distributions are defined based on the MC generated muons, with momenta smeared by the reconstruction resolution. Then, corrections to muon momentum  in bins of $\eta$ and $\phi$ are extracted separately for positively and negatively charged muons. These corrections match the mean values of reconstructed $1/\pt$ spectra to the corresponding reference values. Finally, correction factors are tuned further by comparing  the reconstructed dimuon invariant mass spectra in each $\PGm^+$  and $\PGm^-$ pseudorapidity bin with the reference. The correction factors are determined separately for data and simulated events following the same procedure.

{\tolerance=5000
The overall muon selection efficiency includes contributions from reconstruction, identification, isolation, and trigger efficiencies.
Each component is measured from $\cPZ/\gamma^*\to\Pgmp\Pgmm$ events using the ``tag-and-probe'' method ~\cite{CMS:WZ, CMS-PAPERS-MUO-10-004}.
The efficiencies are measured in bins of $\eta$ and $\pt$ for $\PGm^+$ and $\PGm^-$ separately.
Each $\eta$ bin of the efficiency measurement is fully contained in a single $\etaabs$ bin used for the asymmetry measurement.
The total average efficiency is about 85\% at central rapidities and drops to about 50\% in the last \etaabs bin.
The ratio of the average $\PGmp$ and $\PGmm$ efficiencies varies within 0.6\% of unity in the first 10 $\etaabs$ bins. In the last bin the ratio is 0.98.
The same procedure is used in data and MC simulation, and scale factors are determined to match the MC simulation efficiencies to data.
}

Template shapes, used in the fits, are based on the missing transverse momentum ($\ptvecmiss$) reconstructed with the particle-flow algorithm~\cite{pflow1,pflow2}. The $\ptvecmiss$ is defined as the projection on the plane perpendicular to the beams of the negative vector sum of the momenta of all reconstructed particles in an event. A set of corrections is applied to $\ptvecmiss$ in order to improve the modeling of distributions of $\ETslash=\abs{\ptvecmiss}$ in data and MC templates.  First, the average bias in the $\ptvecmiss$-component along the direction of $\ptvec$-sum of charged particles associated with the pileup vertices is removed~\cite{t0met}. Second, the muon momentum correction to $\ptvec$, described above, is added vectorially to $\ptvecmiss$. In addition, the ``$\phi$-modulation'' corrections, which increase  linearly as a function of pileup, make the $\phi (\ptvecmiss)$ distributions uniform~\cite{t0met}. The above corrections are applied to both data and simulated events. The final set of corrections, derived from the ``hadronic recoil'' technique~\cite{Abazov:2009tra,  cmsmet}, is applied to simulated $\Wpm\to\PGm^{\pm}\PGn$, $\zmumu$, and QCD events to match the average $\ptvecmiss$ scale and resolution to data.

The modeling of the multijet events is further improved with a set of corrections derived from a QCD control sample selected by inverting the offline isolation requirement for events collected using a prescaled muon trigger with no isolation requirement. Muon $\pt$-dependent weight factors are determined for the QCD simulation that match the muon $\pt$ distributions with data. The QCD control sample is also used to derive ratios between the yields with positive and negative muons in each muon $\etaabs$ bin. These ratios are used to constrain the relative QCD contributions to $\PWp$ and $\PWm$ events, as described in Section~\ref{section:signalextraction}.

\section{Signal extraction \label{section:signalextraction}}

In each of the 11 muon $\abs{\eta}$ bins, yields of $\PWp$ and $\PWm$ events are obtained from the simultaneous $\chi^2$-fit of the $\ETslash$ distributions of $\PGm^+$ and $\PGm^-$ events.
The definition of $\chi^2$ used in the fit takes into account the statistical uncertainties in the simulated templates.
The shapes of the $\ETslash$ distributions for the $\Wpm\to\PGm^{\pm}\PGn$ signal and the backgrounds are taken from the MC simulation after correcting for mismodeling of the detector response and for the \pt distribution of $\PW$ bosons.
All electroweak and $\ttbar$ background samples are normalized to the integrated luminosity using the theoretical cross sections calculated at NNLO.
Each simulated event is also weighted with scale factors to match the average muon selection efficiencies in data.
In addition, mass-dependent correction factors are applied to $\zmumu$ simulated events to match the observed mass distribution of dimuon events in data.

The $\PWp$ and $\PWm$ signal yields and the total QCD background normalization are free parameters in each fit.
The relative contributions of QCD background events in the $\PWp$ and $\PWm$ samples are constrained to values obtained from the QCD control sample.
The $\Wpm\to\tau^{\pm}\PGn$ background is normalized to the $\Wpm\to\PGm^{\pm}\PGn$ signal, for each charge, using the scale factors corresponding to the free parameters of the signal yield.
The normalizations of the remaining electroweak and $\ttbar$ backgrounds are fixed in the fit.

\begin{figure*}[hbtp]
  \begin{center}
    \includegraphics[width=\cmsFigWidthNarrow]{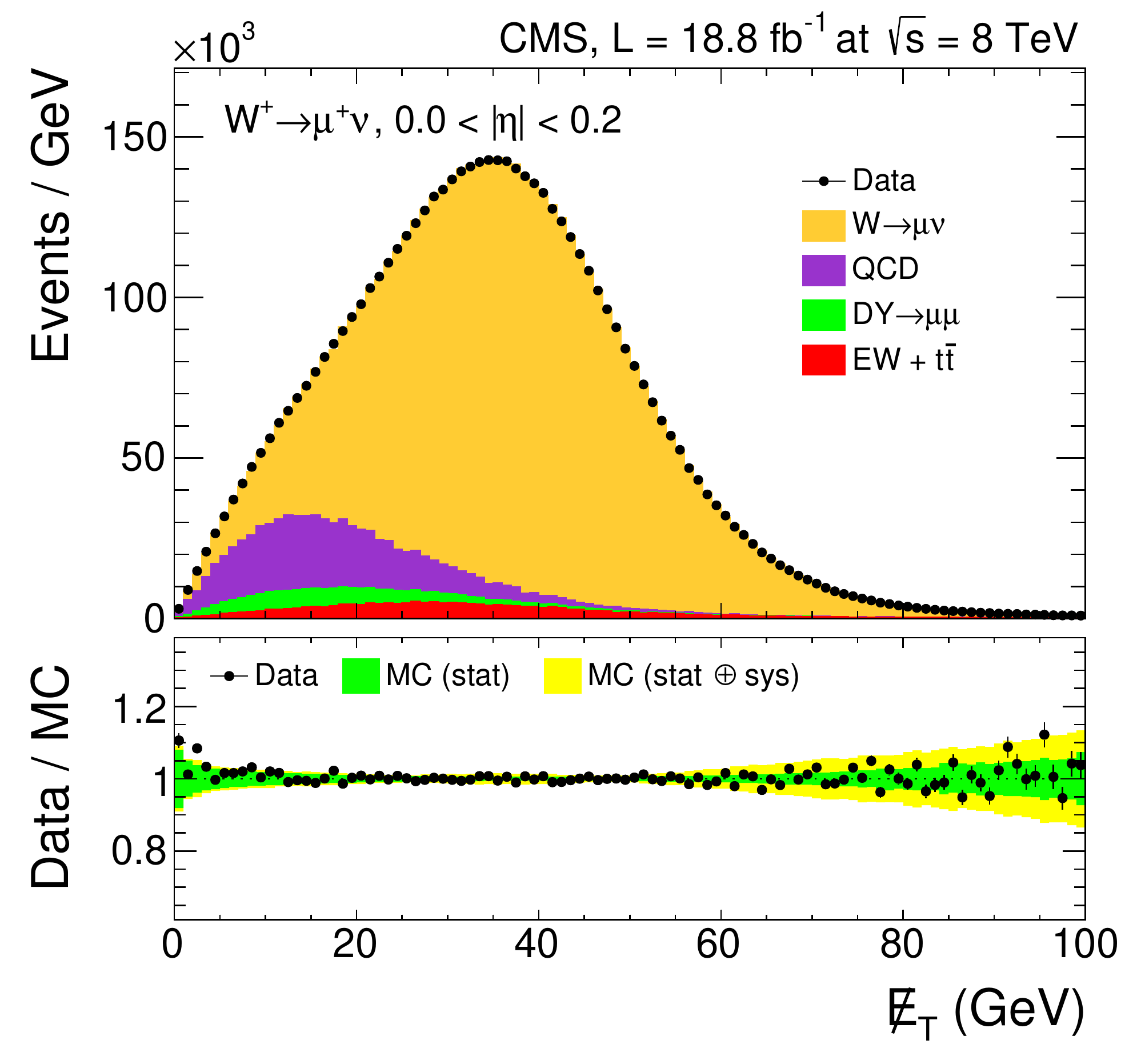}
    \includegraphics[width=\cmsFigWidthNarrow]{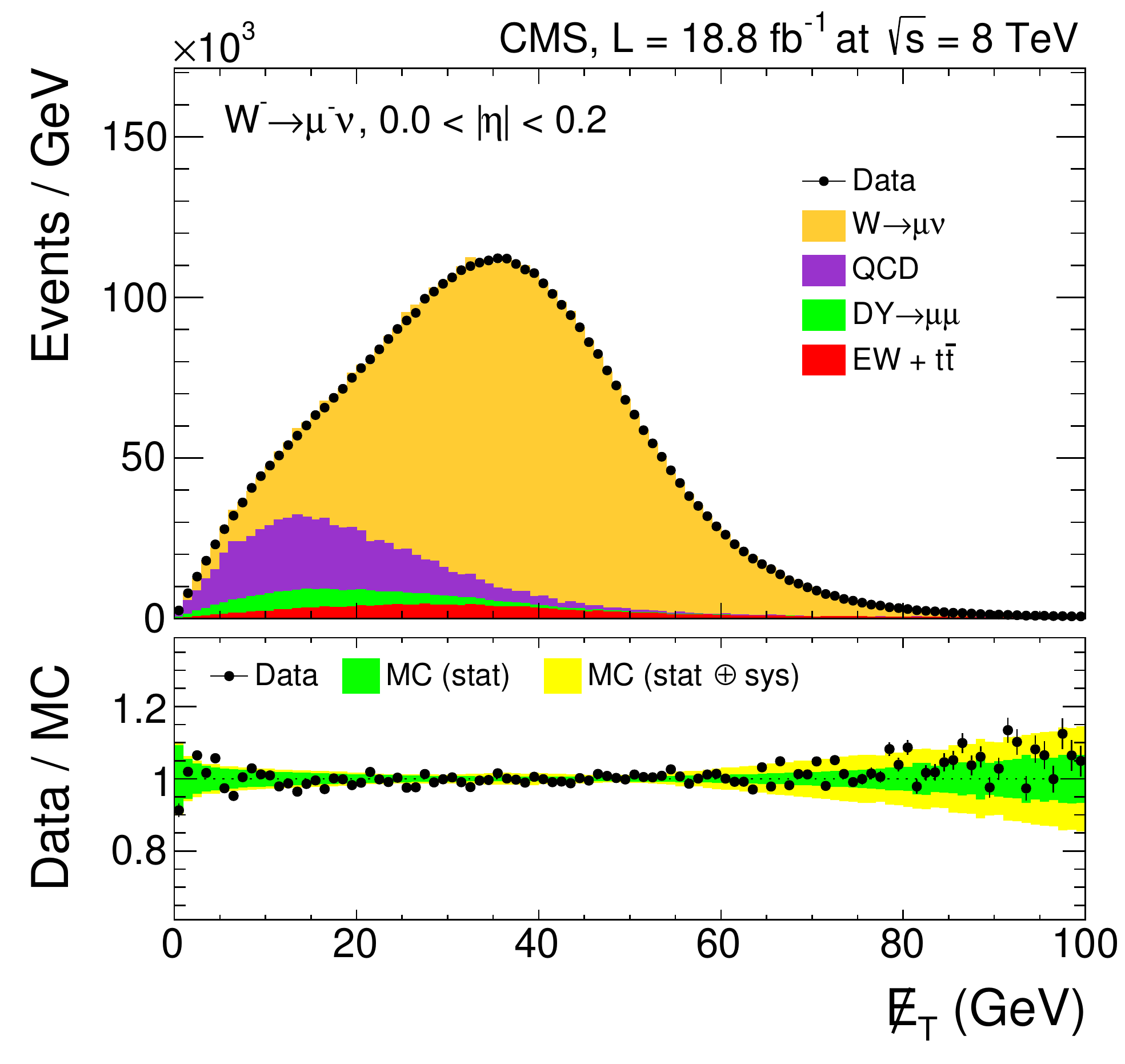}
    \includegraphics[width=\cmsFigWidthNarrow]{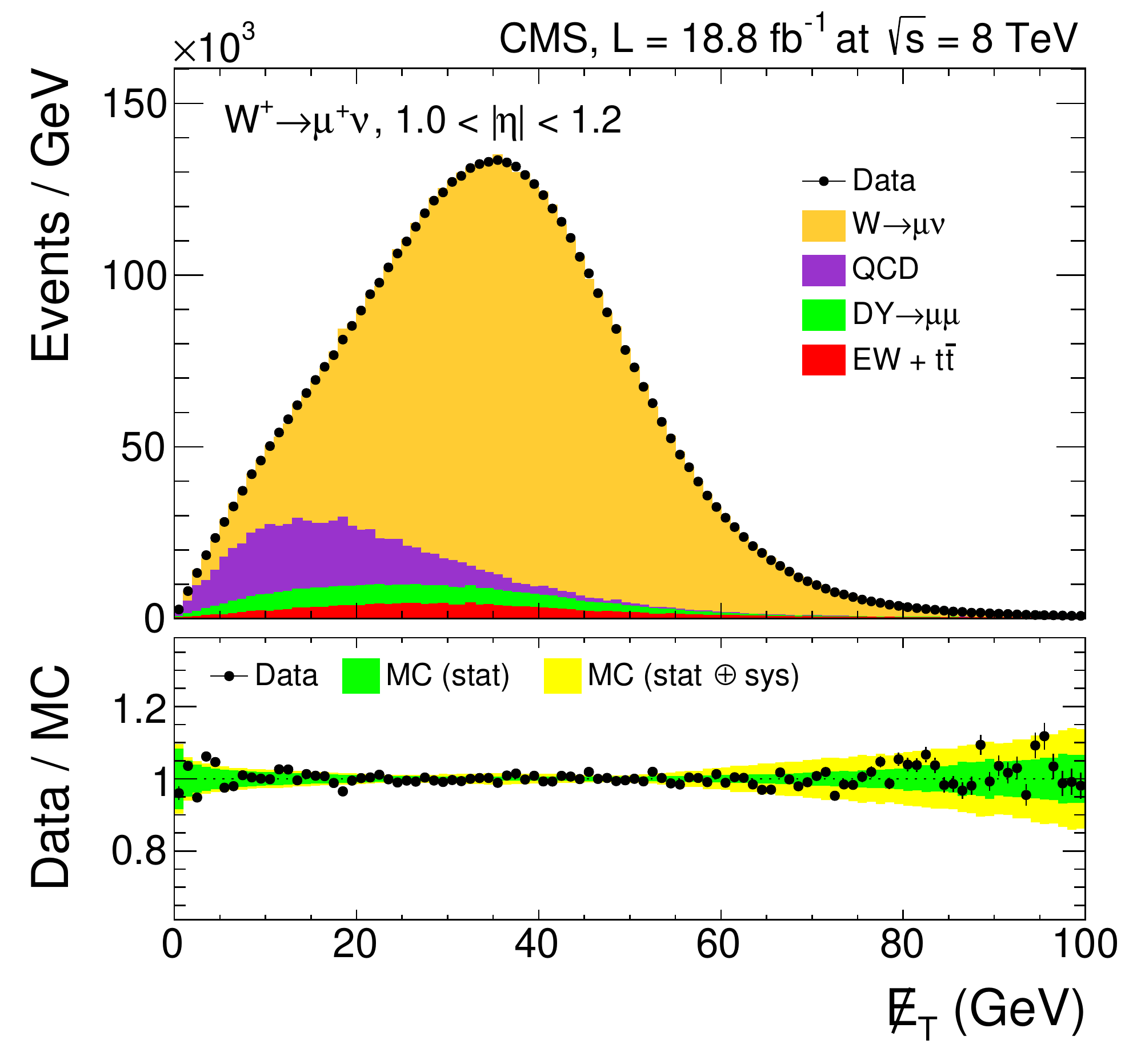}
    \includegraphics[width=\cmsFigWidthNarrow]{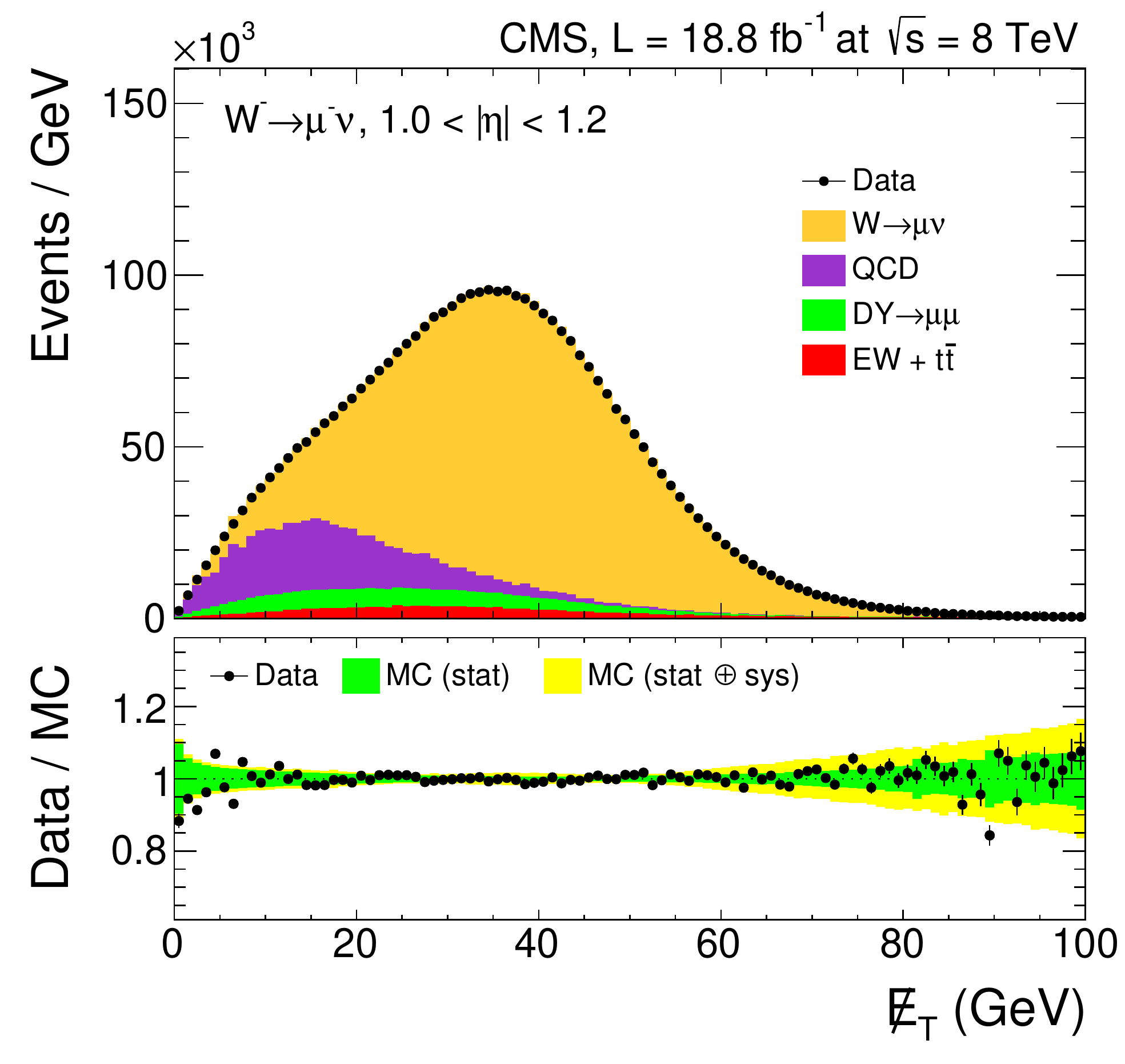}
    \includegraphics[width=\cmsFigWidthNarrow]{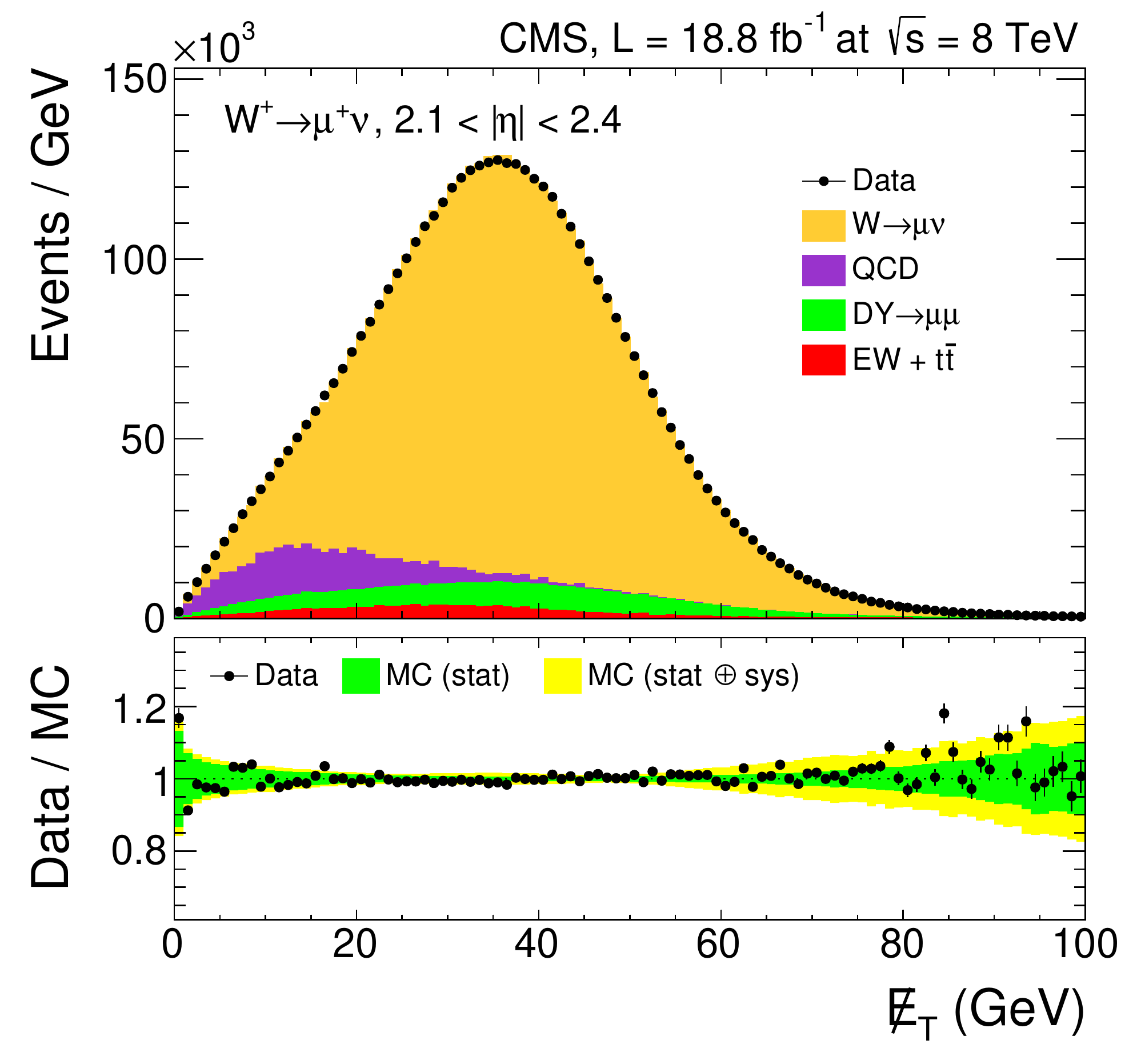}
    \includegraphics[width=\cmsFigWidthNarrow]{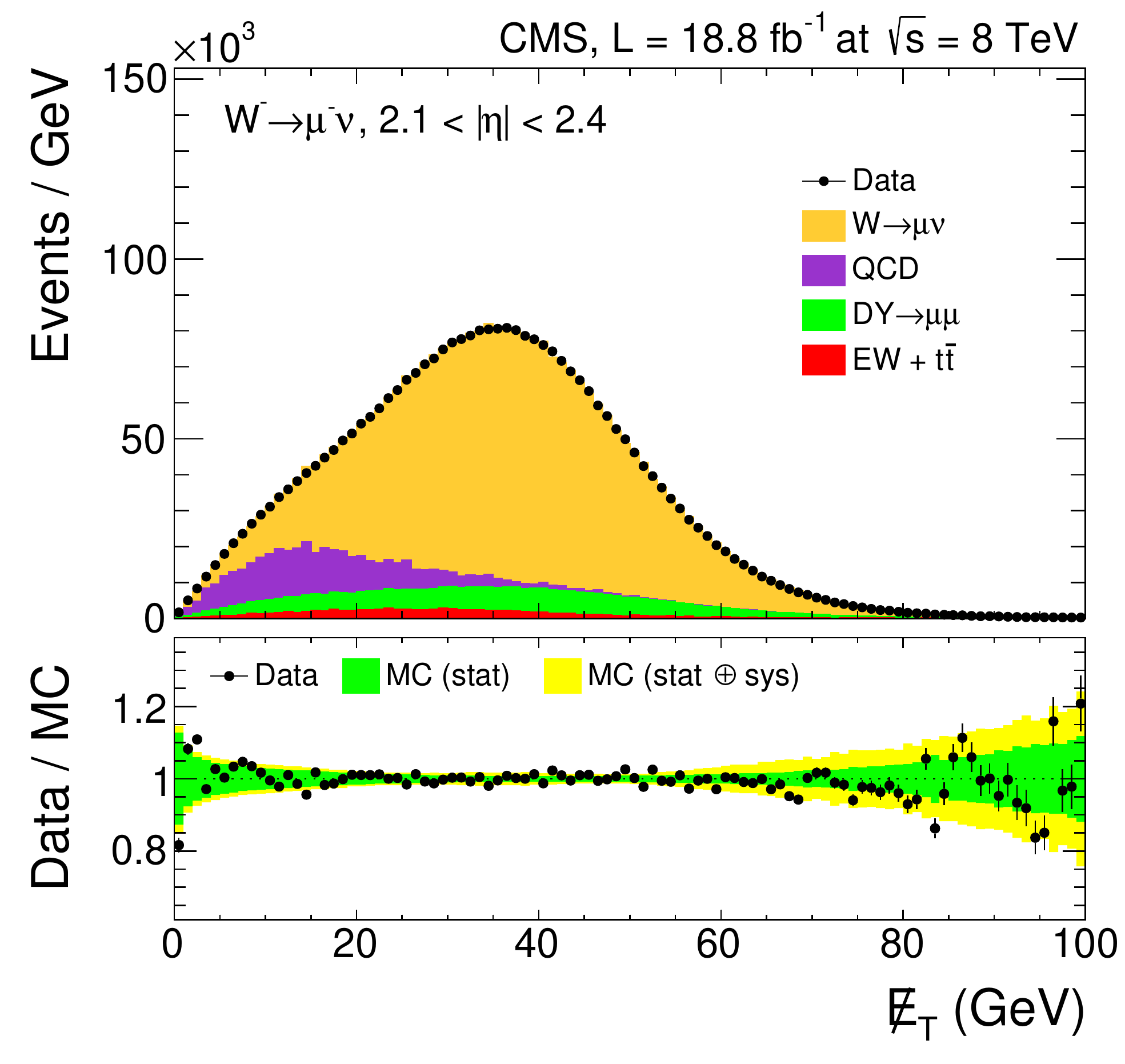}
    \caption{
	Examples of fits in three $\etaabs$ ranges: $0.0<\etaabs<0.2$ (top), $1.0<\etaabs<1.2$ (center), and $2.1<\etaabs<2.4$ (bottom).
	For each $\eta$ range, results for $\PWp$ (left) and $\PWm$ (right) are shown.
	The ratios between the data points  and the final fits are shown at the bottom of each panel.
    }
    \label{fig:results:fits}
  \end{center}
\end{figure*}

Table~\ref{table:fitresults} summarizes the fitted yields of $\PWp$ ($N^{+}$) and $\PWm$ ($N^{-}$) events, the correlation coefficient ($\rho_{+,-}$), and the $\chi^2$ value for each fit. Examples of fits for three $\abs{\eta}$ ranges are shown in Fig.~\ref{fig:results:fits}.
The ratio of the data to the final fit, shown below each distribution,  demonstrates good agreement of the fits with data.
It should be noted that the $\chi^2$ values reported in Table~\ref{table:fitresults} are calculated using the statistical uncertainties of both data and simulated templates; systematic uncertainties are not taken into account.

\begin{table*}[htpb]
\centering
\topcaption{
    Summary of the fitted $N^{+}$, $N^{-}$, the correlation ($\rho_{+,-}$) between the uncertainties in $N^{+}$ and  $N^{-}$,
    and the $\chi^{2}$ of the fit for each $\abs{\eta}$ bin.
    The number of degrees of freedom~($n_{\text{dof}}$) in each fit is 197.
    The quoted uncertainties are statistical and include statistical uncertainties in the templates.
    The correlation coefficients are expressed as percentages.
}
\label{table:fitresults}
\begin{tabular}{ l | c | c c c  }
\hline
$\abs{\eta}$ bin  & $\chi^{2}$ ($n_{\text{dof}}$ = 197) & $N^{+}$~($10^3$)	&   $N^{-}$~($10^3$)   &  $\rho_{+,-}$~(\%)   \\
\hline
0.00--0.20  &  238 & $4648.5\pm4.2$ & $3584.9\pm3.8$ & 18.9  \\
0.20--0.40  &  242 & $4414.5\pm4.0$ & $3360.9\pm3.7$ & 18.8  \\
0.40--0.60  &  248 & $4893.8\pm4.3$ & $3692.5\pm3.9$ & 18.9  \\
0.60--0.80  &  199 & $4900.1\pm4.3$ & $3621.3\pm3.8$ & 19.2  \\
0.80--1.00  &  218 & $4420.8\pm4.0$ & $3218.0\pm3.6$ & 18.7  \\
1.00--1.20  &  204 & $4235.7\pm3.9$ & $2949.2\pm3.4$ & 18.5  \\
1.20--1.40  &  193 & $4176.8\pm3.9$ & $2827.0\pm3.5$ & 19.3  \\
1.40--1.60  &  213 & $4351.2\pm4.2$ & $2864.7\pm3.7$ & 19.3  \\
1.60--1.85  &  208 & $4956.2\pm4.4$ & $3134.1\pm3.9$ & 19.5  \\
1.85--2.10  &  238 & $5292.9\pm4.4$ & $3229.6\pm3.8$ & 18.5  \\
2.10--2.40  &  229 & $4023.7\pm3.9$ & $2428.2\pm3.3$ & 17.6  \\
\hline
\end{tabular}
\end{table*}

For each muon charge and $\abs{\eta}$ bin, the fiducial cross section is calculated as
\begin{equation} \label{eq:xsec}
    \sigeta^{\pm} = \frac{1}{2\Delta\eta}\frac{N^{\pm}}{\epsilon^{\pm}\epsilon_{\mathrm{FSR}}^{\pm}\mathcal{L}_{\mathrm{int}}},
\end{equation}
where $\epsilon^{\pm}$ is the average $\PGm^{\pm}$ selection efficiency per $\abs{\eta}$ bin, $\epsilon_{\mathrm{FSR}}$ takes into account the event loss within the muon $\pt$ acceptance due to the final-state photon emission, and $\mathcal{L}_{\mathrm{int}}$ is the integrated luminosity of the data sample.
Each $\epsilon^{\pm}_{\mathrm{FSR}}$ factor, defined as a ratio of the numbers of events within the $\pt$ acceptance after and before FSR, is evaluated using the signal MC samples.

\section{Systematic uncertainties}

To estimate the systematic uncertainties in the muon selection efficiencies, several variations are applied to the measured efficiency tables. First, the efficiency values in each $\eta$--$\pt$ bin are varied within their statistical errors for data and simulation independently. In each pseudo-experiment the varied set of efficiencies is used to correct the MC simulation templates and extract the cross sections using Eq.~(\ref{eq:xsec}). The standard deviation of the resulting distribution is taken as a systematic uncertainty for each charge and $\etaabs$ bin. The statistical uncertainties between the two charges and all $\eta$--$\pt$ bins are uncorrelated. Second, the offline muon selection efficiency scale factors are varied by ${\pm}0.5\%$ coherently for both charges and all bins. The trigger selection efficiency scale factors are varied by ${\pm}0.2\%$, assuming no correlations between the $\eta$ bins, but ${+}100\%$ correlations between the charges and $\pt$ bins. The above systematic variations take into account uncertainties associated with the tag-and-probe technique and are assessed by varying signal and background dimuon mass shapes, levels of background, and dimuon mass range and binning used in the fits. Finally, an additional ${\pm}100\%$ correlated variation is applied based on the bin-by-bin difference between the true and measured efficiencies in the $\zmumu$ MC sample. This difference changes gradually from about 0.5\% in the first bin to about ${-}2\%$ in the last bin. This contribution is the main source of negative correlations in the systematic uncertainties between the central and high rapidity bins.
The total systematic uncertainty in the efficiency is obtained by adding up the four covariance matrices corresponding to the above variations.

A possible mismeasurement of the charge of the muon could lead to a bias in the observed asymmetry between the $\PWp$ and $\PWm$ event rates. The muon charge misidentification rate has been studied in detail and was found to be negligible ($10^{-5}$)~\cite{CMS:asym:2010}.

The muon momentum correction affects the yields and the shapes of the $\ETslash$ distributions in both data and MC simulation. To estimate the systematic uncertainty, the muon correction parameters in each $\eta$--$\phi$ bin and overall scale are varied within their uncertainties. The standard deviation of the resulting cross section distribution for each charge and muon $\etaabs$ bin is taken as the systematic uncertainty and the corresponding correlations are calculated. Finite detector resolution effects, which result in the migration of events around the $\pt$ threshold and between $\etaabs$ bins, have been studied with the signal MC sample and found to have a negligible impact on the measured cross sections and asymmetries.

\begin{figure}[!htbp]
\center
   \includegraphics[width=\cmsFigWidth]{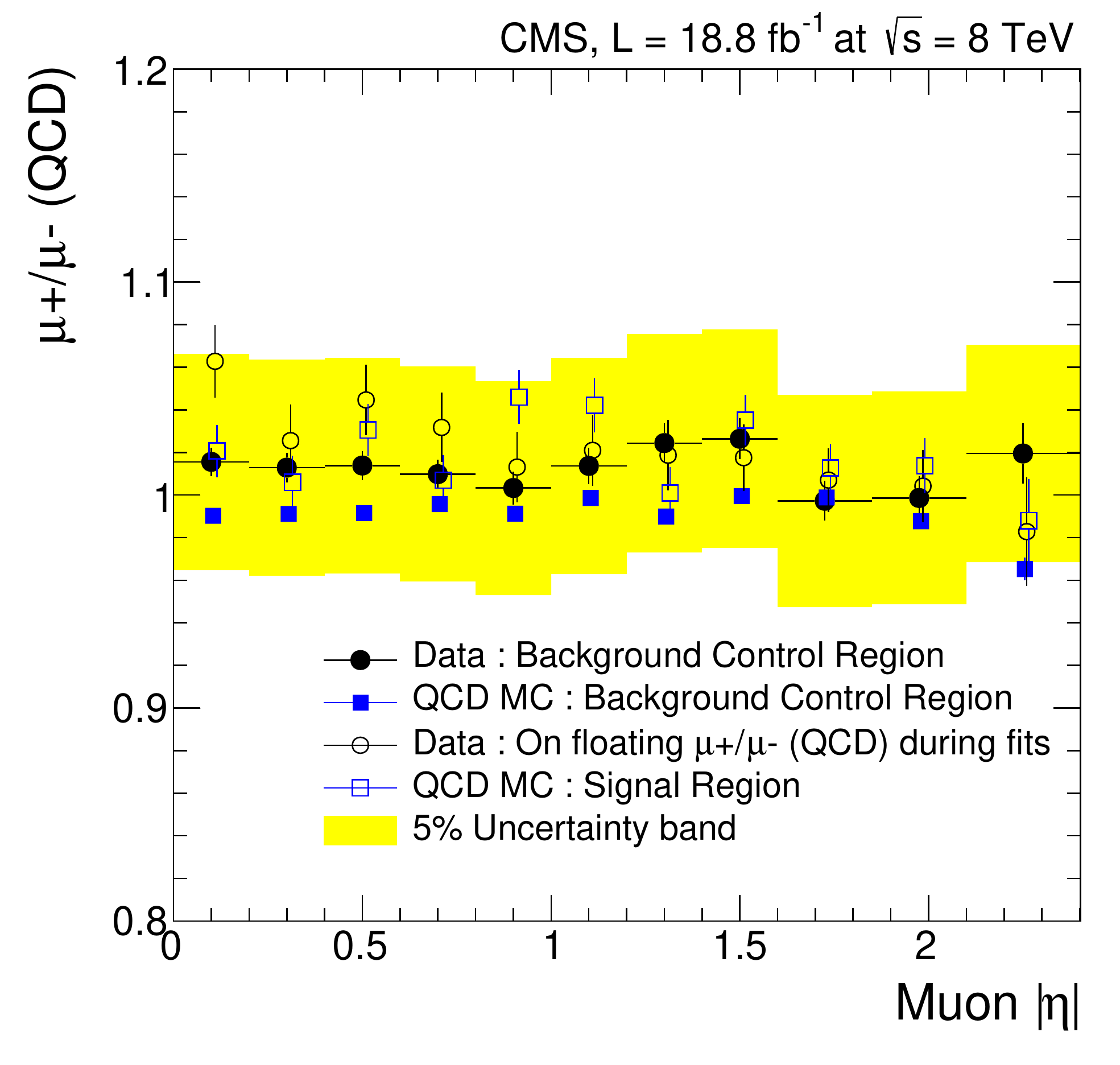}
\caption{
Distribution of \RQCD in QCD control region for data (solid circles),  QCD control region for simulation (solid squares), and signal region for simulation (open squares).
Open circles show the \RQCD distribution when QCD contributions in $\PWp$ and $\PWm$ events are not constrained. Shaded area indicates assigned systematic uncertainty.
}
\label{fig:qcdpm}
\end{figure}

There are two sources of systematic uncertainties associated with the QCD background estimate.
One is the uncertainty in the ratio of QCD background events in the $\PWp$ and $\PWm$ samples~(\RQCD).
Whereas the total QCD normalization is one of the free parameters in the fit, \RQCD is constrained to the value observed in the QCD control sample, which varies within 3\% of unity depending on the $\etaabs$ bin.
The corresponding systematic uncertainty is evaluated by changing it by ${\pm}5\%$ in each $\etaabs$ bin.
This variation covers the maximum deviations indicated by the QCD MC simulation, as indicated in Fig.~\ref{fig:qcdpm}.
The resulting systematic uncertainties are assumed to be uncorrelated between the $\etaabs$ bins.
Additionally, to take into account possible bias in this ratio due to different flavor composition in the signal and QCD control regions,
the average difference of this ratio between the signal and QCD control regions is evaluated using the QCD MC simulation.
This difference of about 3\% is taken as an additional 100\%-correlated systematic uncertainty in \RQCD.
As a check, using the same shape for the QCD background in  $\Pgmp$ and $\Pgmm$ events, its normalization is allowed to float independently for the two charges.
The resulting values of \RQCD are covered by the above systematic uncertainties.

The other component of systematic uncertainty, associated with the QCD background, is the $\ETslash$ shape.
To estimate the systematic uncertainty in modeling the shape of the $\ETslash$ distributions in QCD events, additional fits are performed using the $\ETslash$ distributions without the hadronic recoil corrections and the $\pt$-dependent scale factors; this results in a variation of about 2\% in the average $\ETslash$ resolution.
The resulting shifts in the extracted cross section values in each \etaabs bin are taken as systematic uncertainties. The correlations between the $\etaabs$ bins and the two charges are assumed to be 100\%.

The normalization of the $\zmumu$ background in the signal region is corrected with mass-dependent scale factors that match the dimuon mass distribution in MC simulation with data. The systematic uncertainty is the difference between the cross sections calculated with and without applying these corrections to the DY background normalizations. An uncertainty of 7\% is assigned to the $\ttbar$ theoretical cross section~\cite{toppp}, which is used to normalize the $\ttbar$ background to the integrated luminosity of the data sample. In the fits, the $\Wpm\to\tau^{\pm}\PGn$ background is normalized to the $\Wpm\to\PGm^{\pm}\PGn$ yields in data with a ratio obtained from the simulation. A ${\pm}2\%$ uncertainty is assigned to the $\Wpm\to\tau^{\pm}\PGn$ to $\Wpm\to\PGm^{\pm}\PGn$ ratio~\cite{Agashe:2014kda}. Each of the above variations is assumed to be fully correlated for the different bins and the two charges.

There are several sources of systematic uncertainty that affect the $\ETslash$ shapes. The systematic uncertainty associated with the $\phi$ modulation of $\ptvecmiss$ is small and is evaluated by removing the corresponding correction to $\ptvecmiss$.
A 5\% uncertainty is assigned to the minimum bias cross section used to calculate the expected pileup distribution in data.
To improve the agreement between data and simulation, the $\PW$ boson $\pt$ spectrum is weighted using factors determined by the ratios of the \pt distributions in  $\zmumu$ events in data and MC simulation. The difference in measured cross sections with and without this correction is taken as a systematic uncertainty. Each of the sources above are assumed to be fully correlated between the two charges and different bins. Systematic uncertainties associated with the recoil corrections are evaluated by varying the average recoil and resolution parameters within their uncertainties. The standard deviation of the resulting cross section distribution is taken as the systematic uncertainty and correlations between the two charges and different bins are calculated.

The emission of FSR photons tends, on average, to reduce the muon $\pt$. The observed post-FSR cross sections within the $\pt$ acceptance are corrected using the $\epsilon_{\mathrm{FSR}}^\pm$ factors derived from the signal MC sample. The difference between the $\pt$ spectra of positive and negative muons results in smaller charge asymmetries after FSR compared with those before FSR within the same $\pt>25\GeV$ acceptance. These differences, which vary between 0.07\% and 0.11\% depending on the $\etaabs$ bin, are corrected by the charge-dependent $\epsilon_{\mathrm{FSR}}^\pm$ efficiency factors. The systematic uncertainty in the FSR modeling is estimated by reweighting events with radiated photons with correction factors that account for missing electroweak corrections in the parton shower~\cite{photos,Burkhardt:2001xp}. The $\epsilon^{\pm}_{\mathrm{FSR}}$ correction factors are reevaluated after such reweighting, and the difference between the cross sections, calculated with the new and default $\epsilon_{\mathrm{FSR}}^\pm$ values, is taken as a systematic uncertainty. The correlations between the two charges and different $\etaabs$ bins are assumed to be $100\%$. The effects of migration between the $\etaabs$ bins due to final-state photon emission have been evaluated with the signal MC sample and are found to be negligible.

The PDF uncertainties are evaluated by using the NLO MSTW2008~\cite{Martin:2009ad}, {CT10}~\cite{CTEQ:1007}, and NN\-PDF2.1~\cite{Ball:2011mu} PDF sets.
All simulated events are weighted according to a given PDF set, varying both the template normalizations and shapes.
For CT10 and MSTW2008  PDFs, asymmetric master equations are used~\cite{Martin:2009ad, CTEQ:1007}.
For the NNPDF2.1 PDF set, the standard deviation of the extracted cross section distributions is taken as a systematic uncertainty.
For the CT10, the 90\% confidence level (\CL) uncertainty is rescaled to 68\% \CL using a factor of 1.645.
The half-width of the total envelope of all three PDF uncertainty bands is taken as the PDF uncertainty.
The CT10 error set is used to estimate the correlations between the two charges and different $\etaabs$ bins.

Finally, a ${\pm}2.6\%$ uncertainty~\cite{CMS-PAS-LUM-13-001} is assigned to the integrated luminosity of the data sample.
The luminosity uncertainty is fully correlated between the $\etaabs$ bins and two charges. Therefore, this uncertainty cancels in the measured charge asymmetries.
The uncertainty in the normalization of the electroweak backgrounds due to the luminosity uncertainty has a negligible impact on the measurements.

\begin{table*}[t]
\centering
\topcaption{
    Systematic uncertainties in cross sections ($\delta\sigeta^\pm$) and charge asymmetry ($\delta\mathcal{A}$) for each $\abs{\eta}$ bin.
    The statistical and integrated luminosity uncertainties are also shown for comparison.
    A detailed description of each systematic uncertainty is given in the text.
}
\label{table:systematics}
\resizebox{\textwidth}{!}{
\begin{tabular}{ l | r  r  r  r  r  r  r  r  r  r  r  }
\hline
$\abs{\eta}$ bin   &   0.0--0.2 & 0.2--0.4 & 0.4--0.6 & 0.6--0.8 & 0.8--1.0 & 1.0--1.2 & 1.2--1.4 & 1.4--1.6 & 1.6--1.85 & 1.85--2.1 & 2.1--2.4 \\
\hline
\hline
\multicolumn{12}{c}{$\delta\sigeta^+$~(pb)} \\
\hline
Efficiency		 &   5.5 &   7.0 &   6.3 &   6.2 &   6.6 &   4.5 &   4.3 &   4.3 &   5.3 &   6.9 &  17.7 \\ 	
Muon scale		 &   0.4 &   0.4 &   0.4 &   0.4 &   0.4 &   0.5 &   0.5 &   0.5 &   0.5 &   0.5 &   0.4 \\
QCD  $+/-$		 &   1.1 &   1.1 &   1.1 &   1.1 &   1.1 &   1.1 &   1.2 &   1.3 &   1.2 &   1.0 &   0.7 \\
QCD shape		 &   2.0 &   1.9 &   2.0 &   2.0 &   1.9 &   1.9 &   2.0 &   2.3 &   2.1 &   1.5 &   1.0 \\ 	
EW+$\ttbar$ bkg	         &   0.6 &   0.6 &   0.6 &   0.6 &   0.6 &   0.7 &   0.8 &   0.9 &   1.0 &   1.2 &   1.3 \\
$\ETslash$ shape	 &   2.4 &   2.4 &   2.5 &   2.4 &   2.4 &   2.5 &   2.8 &   2.9 &   2.9 &   2.4 &   1.9 \\ 	
PDF			 &   0.6 &   0.5 &   0.5 &   0.5 &   0.7 &   0.9 &   1.1 &   1.3 &   1.6 &   1.8 &   2.0 \\
FSR			 &   0.2 &   0.2 &   0.2 &   0.2 &   0.2 &   0.2 &   0.2 &   0.1 &   0.1 &   0.0 &   0.0 \\ \hline
Total syst.		 &   6.5 &   7.7 &   7.2 &   7.1 &   7.4 &   5.8 &   5.8 &   6.1 &   6.8 &   7.9 &  18.0 \\ \hline
Int. lum.                &  19.3 &  19.5 &  19.5 &  19.6 &  19.8 &  19.9 &  20.1 &  20.1 &  20.2 &  20.0 &  19.5 \\ \hline
Stat.			 &   0.7 &   0.7 &   0.7 &   0.7 &   0.7 &   0.7 &   0.7 &   0.7 &   0.7 &   0.6 &   0.7 \\ \hline
Total unc.		 &  20.4 &  21.0 &  20.8 &  20.9 &  21.2 &  20.7 &  21.0 &  21.0 &  21.3 &  21.6 &  26.5 \\ \hline
\hline
\multicolumn{12}{c}{$\delta\sigeta^-$~(pb)} \\
\hline
Efficiency		 &   4.3 &   5.5 &   4.8 &   4.6 &   4.4 &   3.2 &   2.9 &   2.8 &   3.4 &   4.0 &  10.1 \\ 	
Muon scale		 &   0.3 &   0.3 &   0.3 &   0.3 &   0.3 &   0.3 &   0.3 &   0.4 &   0.4 &   0.3 &   0.3 \\
QCD  $+/-$		 &   0.9 &   0.9 &   0.9 &   0.9 &   0.9 &   0.9 &   1.0 &   1.0 &   1.0 &   0.8 &   0.5 \\
QCD shape		 &   1.8 &   1.8 &   1.9 &   1.8 &   1.8 &   1.8 &   2.0 &   2.1 &   2.1 &   1.5 &   1.0 \\ 	
EW+$\ttbar$ bkg		 &   0.5 &   0.5 &   0.5 &   0.6 &   0.6 &   0.6 &   0.7 &   0.8 &   0.9 &   1.1 &   1.2 \\
$\ETslash$ shape	 &   2.2 &   2.2 &   2.3 &   2.2 &   2.3 &   2.4 &   2.5 &   2.8 &   2.7 &   2.3 &   1.7 \\ 	
PDF			 &   0.6 &   0.6 &   0.5 &   0.5 &   0.7 &   0.8 &   1.1 &   1.2 &   1.5 &   1.6 &   1.9 \\
FSR			 &   0.2 &   0.2 &   0.1 &   0.1 &   0.1 &   0.1 &   0.1 &   0.1 &   0.1 &   0.1 &   0.1 \\ \hline
Total syst.		 &   5.3 &   6.4 &   5.7 &   5.5 &   5.5 &   4.6 &   4.6 &   4.9 &   5.2 &   5.3 &  10.6 \\ \hline
Int. lum.		 &  14.8 &  14.8 &  14.7 &  14.5 &  14.3 &  13.9 &  13.6 &  13.2 &  12.7 &  12.1 &  11.4 \\ \hline
Stat.			 &   0.6 &   0.6 &   0.6 &   0.6 &   0.6 &   0.6 &   0.6 &   0.6 &   0.6 &   0.6 &   0.6 \\ \hline
Total unc.		 &  15.7 &  16.1 &  15.8 &  15.5 &  15.3 &  14.7 &  14.3 &  14.1 &  13.7 &  13.2 &  15.6 \\ \hline
\hline
\multicolumn{12}{c}{$\delta\mathcal{A}\times100$} \\
\hline
Efficiency		 &   0.06 &   0.07 &   0.06 &   0.06 &   0.09 &   0.09 &   0.10 &   0.09 &   0.09 &   0.08 &   0.14 \\ 	
Muon scale		 &   0.03 &   0.03 &   0.03 &   0.03 &   0.03 &   0.03 &   0.03 &   0.04 &   0.04 &   0.03 &   0.03 \\
QCD  $+/-$		 &   0.15 &   0.15 &   0.15 &   0.15 &   0.15 &   0.15 &   0.16 &   0.18 &   0.17 &   0.14 &   0.10 \\
QCD shape		 &   0.02 &   0.03 &   0.03 &   0.03 &   0.04 &   0.04 &   0.05 &   0.06 &   0.07 &   0.05 &   0.04 \\ 	
EW+$\ttbar$ bkg 	 &   0.03 &   0.03 &   0.03 &   0.03 &   0.03 &   0.03 &   0.03 &   0.03 &   0.04 &   0.04 &   0.05 \\
$\ETslash$ shape	 &   0.03 &   0.04 &   0.04 &   0.04 &   0.05 &   0.06 &   0.06 &   0.09 &   0.09 &   0.09 &   0.07 \\ 	
PDF			 &   0.03 &   0.02 &   0.02 &   0.02 &   0.02 &   0.03 &   0.04 &   0.05 &   0.05 &   0.09 &   0.08 \\
FSR			 &   0.00 &   0.00 &   0.00 &   0.00 &   0.00 &   0.00 &   0.00 &   0.00 &   0.00 &   0.00 &   0.01 \\ \hline
Total syst.		 &   0.17 &   0.18 &   0.18 &   0.18 &   0.19 &   0.20 &   0.22 &   0.23 &   0.23 &   0.22 &   0.21 \\ \hline
Stat.			 &   0.06 &   0.06 &   0.06 &   0.06 &   0.06 &   0.07 &   0.07 &   0.07 &   0.06 &   0.06 &   0.07 \\ \hline
Total unc.		 &   0.18 &   0.19 &   0.19 &   0.19 &   0.20 &   0.21 &   0.23 &   0.24 &   0.24 &   0.23 &   0.22 \\ \hline

\end{tabular}
}
\end{table*}

\begin{table*}[htbp]
\centering
\topcaption{
Correlation matrices of systematic uncertainties for $\sigeta^\pm$ and $\mathcal{A}$.
The statistical and integrated luminosity uncertainties are not included.
The full $22\times22$ correlation matrix for $\sigeta^\pm$ is presented as four blocks of $11\times11$ matrices, as shown in Eq.~(\ref{eq:covariance}).
The $C_{++}$ and $C_{--}$ blocks on the diagonal represent the bin-to-bin correlations of $\delta\sigeta^+$ and $\delta\sigeta^-$, respectively.
The off-diagonal $C_{+-}$ and $C_{+-}^T$ blocks describe the correlations between the two charges.
The values are expressed as percentages.}

\label{table:correlation}
\resizebox{\textwidth}{!}{
\begin{tabular}{ l | r  r  r  r  r  r  r  r  r  r  r }
\hline
$\abs{\eta}$ bin   &   0.0--0.2 & 0.2--0.4 & 0.4--0.6 & 0.6--0.8 & 0.8--1.0 & 1.0--1.2 & 1.2--1.4 & 1.4--1.6 & 1.6--1.85 & 1.85--2.1 & 2.1--2.4 \\ \hline
\hline
\multicolumn{12}{ c }{Correlation matrix of systematic uncertainties in $\sigeta^\pm$ } \\
\hline
\multicolumn{12}{c}{$C_{++}$} \\
\hline
0.00--0.20 & 100.0 &  90.3 &  91.3 &  91.2 &  90.2 &  81.8 &  69.4 &  63.3 &  32.9 &   6.1 & $-$37.2 \\
0.20--0.40 &       & 100.0 &  92.5 &  92.4 &  92.2 &  74.9 &  58.6 &  51.1 &  16.2 & $-$11.8 & $-$54.7 \\
0.40--0.60 &       &       & 100.0 &  92.5 &  92.1 &  79.2 &  64.8 &  57.9 &  25.0 &  $-$2.8 & $-$46.6 \\
0.60--0.80 &       &       &       & 100.0 &  92.2 &  79.5 &  65.4 &  58.6 &  25.8 &  $-$1.9 & $-$46.0 \\
0.80--1.00 &       &       &       &       & 100.0 &  78.3 &  63.7 &  56.7 &  23.2 &  $-$4.5 & $-$48.7 \\
1.00--1.20 &       &       &       &       &       & 100.0 &  82.8 &  79.9 &  60.0 &  38.5 &  $-$3.9 \\
1.20--1.40 &       &       &       &       &       &       & 100.0 &  85.6 &  74.8 &  58.4 &  20.5 \\
1.40--1.60 &       &       &       &       &       &       &       & 100.0 &  79.8 &  65.6 &  30.0 \\
1.60--1.85 &       &       &       &       &       &       &       &       & 100.0 &  86.6 &  64.5 \\
1.85--2.10 &       &       &       &       &       &       &       &       &       & 100.0 &  83.8 \\
2.10--2.40 &       &       &       &       &       &       &       &       &       &       & 100.0 \\
\hline
\multicolumn{12}{c}{$C_{--}$} \\
\hline
0.00--0.20 & 100.0 &  91.1 &  92.1 &  91.9 &  91.2 &  81.8 &  64.8 &  65.6 &  38.5 &  19.0 & $-$31.5 \\
0.20--0.40 &       & 100.0 &  92.8 &  92.3 &  91.2 &  74.6 &  53.0 &  54.4 &  23.1 &   2.2 & $-$48.8 \\
0.40--0.60 &       &       & 100.0 &  92.8 &  92.1 &  80.1 &  61.2 &  62.5 &  33.3 &  12.8 & $-$38.7 \\
0.60--0.80 &       &       &       & 100.0 &  92.3 &  81.3 &  63.4 &  64.6 &  36.2 &  15.8 & $-$35.9 \\
0.80--1.00 &       &       &       &       & 100.0 &  82.5 &  65.4 &  66.8 &  39.0 &  18.6 & $-$33.3 \\
1.00--1.20 &       &       &       &       &       & 100.0 &  83.5 &  84.1 &  67.9 &  52.2 &   5.4 \\
1.20--1.40 &       &       &       &       &       &       & 100.0 &  88.9 &  83.9 &  73.2 &  35.4 \\
1.40--1.60 &       &       &       &       &       &       &       & 100.0 &  83.8 &  72.6 &  33.1 \\
1.60--1.85 &       &       &       &       &       &       &       &       & 100.0 &  88.5 &  64.4 \\
1.85--2.10 &       &       &       &       &       &       &       &       &       & 100.0 &  80.4 \\
2.10--2.40 &       &       &       &       &       &       &       &       &       &       & 100.0 \\
\hline
\multicolumn{12}{c}{$C_{+-}$} \\
\hline
0.00--0.20 &  92.7 &  89.5 &  90.1 &  89.9 &  89.0 &  78.7 &  61.1 &  61.7 &  34.6 &  16.0 & $-$32.9 \\
0.20--0.40 &  88.9 &  94.4 &  90.7 &  90.1 &  88.8 &  71.2 &  49.0 &  50.1 &  18.8 &  $-$1.2 & $-$50.5 \\
0.40--0.60 &  90.0 &  91.7 &  93.6 &  90.9 &  89.9 &  76.0 &  55.9 &  56.9 &  27.4 &   7.6 & $-$42.2 \\
0.60--0.80 &  89.7 &  91.4 &  90.9 &  93.2 &  89.9 &  76.2 &  56.4 &  57.3 &  28.1 &   8.3 & $-$41.6 \\
0.80--1.00 &  88.8 &  91.3 &  90.5 &  90.4 &  92.0 &  74.9 &  54.6 &  55.7 &  25.9 &   5.9 & $-$44.1 \\
1.00--1.20 &  80.4 &  74.4 &  78.8 &  80.1 &  80.8 &  87.9 &  77.4 &  77.4 &  60.4 &  45.7 &   0.8 \\
1.20--1.40 &  68.5 &  58.5 &  65.4 &  67.3 &  68.7 &  82.1 &  86.3 &  82.8 &  74.5 &  63.8 &  25.2 \\
1.40--1.60 &  62.6 &  51.2 &  59.0 &  61.2 &  62.9 &  80.1 &  84.7 &  86.4 &  79.6 &  70.5 &  34.8 \\
1.60--1.85 &  32.7 &  16.8 &  26.9 &  29.8 &  32.2 &  61.8 &  78.4 &  77.1 &  89.1 &  87.0 &  68.2 \\
1.85--2.10 &   6.0 & $-$11.4 &  $-$0.8 &   2.3 &   4.9 &  40.4 &  63.9 &  61.7 &  82.1 &  91.4 &  86.0 \\
2.10--2.40 & $-$36.9 & $-$54.0 & $-$44.2 & $-$41.7 & $-$39.3 &  $-$1.0 &  29.0 &  26.2 &  58.5 &  75.1 &  98.2 \\
\hline
\hline
\multicolumn{12}{c}{Correlation matrix of systematic uncertainties in $\mathcal{A}$} \\
\hline

0.00--0.20 &100.0 &  27.2 &  27.4 &  26.8 &  24.8 &  26.7 &  24.3 &  27.0 &  26.1 &  25.0 &  19.7 \\
0.20--0.40 &      & 100.0 &  27.8 &  27.2 &  24.5 &  27.3 &  24.1 &  28.9 &  27.8 &  28.1 &  22.5 \\
0.40--0.60 &      &       & 100.0 &  28.3 &  27.4 &  28.6 &  27.0 &  29.9 &  29.5 &  28.0 &  21.4 \\
0.60--0.80 &      &       &       & 100.0 &  29.3 &  29.0 &  29.1 &  30.6 &  30.9 &  28.7 &  22.2 \\
0.80--1.00 &      &       &       &       & 100.0 &  29.7 &  32.8 &  32.0 &  33.0 &  28.4 &  20.7 \\
1.00--1.20 &      &       &       &       &       & 100.0 &  31.2 &  33.6 &  34.3 &  31.9 &  25.4 \\
1.20--1.40 &      &       &       &       &       &       & 100.0 &  34.2 &  36.8 &  30.6 &  25.1 \\
1.40--1.60 &      &       &       &       &       &       &       & 100.0 &  39.1 &  37.4 &  31.0 \\
1.60--1.85 &      &       &       &       &       &       &       &       & 100.0 &  38.6 &  33.3 \\
1.85--2.10 &      &       &       &       &       &       &       &       &       & 100.0 &  42.0 \\
2.10--2.40 &      &       &       &       &       &       &       &       &       &       & 100.0 \\
\hline

\end{tabular}
}
\end{table*}

Table~\ref{table:systematics} summarizes the systematic uncertainties in the measured cross sections and asymmetries.
For comparison, the statistical and luminosity uncertainties are also shown.
The uncertainty in the integrated luminosity dominates the total uncertainties in the measured cross sections, while the uncertainty in the QCD background
estimation dominates the uncertainties in the charge asymmetries. The uncertainties for the muon charge asymmetries are calculated from those in the differential cross sections, taking into account the correlations between the two charges.

The correlations in the systematic uncertainty between the charges and different $\etaabs$ bins are shown in Table~\ref{table:correlation}.
The full $22{\times}22$ correlation matrix $C$ is split into four $11{\times}11$ blocks as
\begin{equation}
\label{eq:covariance}
C=
\begin{bmatrix}
    C_{++}   & C_{+-}  \\
    C_{+-}^T & C_{--}   \\
\end{bmatrix},
\end{equation}
where the $C_{++}$ and $C_{--}$ matrices represent the bin-to-bin correlations of systematic uncertainties in $\sigeta^+$ and $\sigeta^-$, respectively, and $C_{+-}$ describes the correlations between the two charges. To construct the total covariance matrix, the covariance matrix of the systematic uncertainties should be added to those of the statistical and integrated luminosity uncertainties. The latter are fully correlated between the two charges and $\etaabs$ bins. For the statistical uncertainties bin-to-bin correlations are zero; the correlations between the two charges are shown in Table~\ref{table:fitresults}.

\section{Results}

\begin{figure*}[htbp]
\centering
    \includegraphics[width=\cmsFigWidthNarrow]{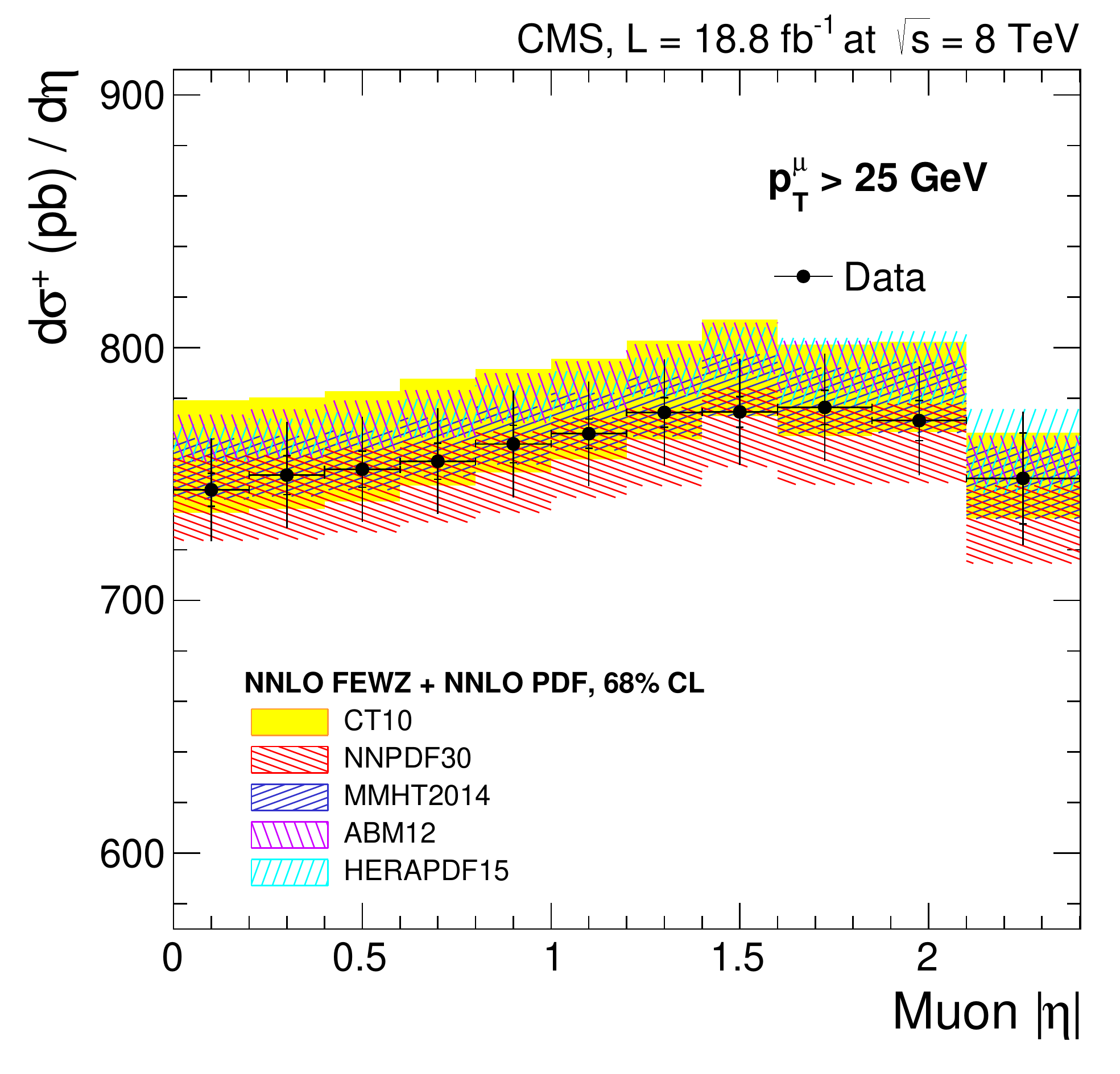}
    \includegraphics[width=\cmsFigWidthNarrow]{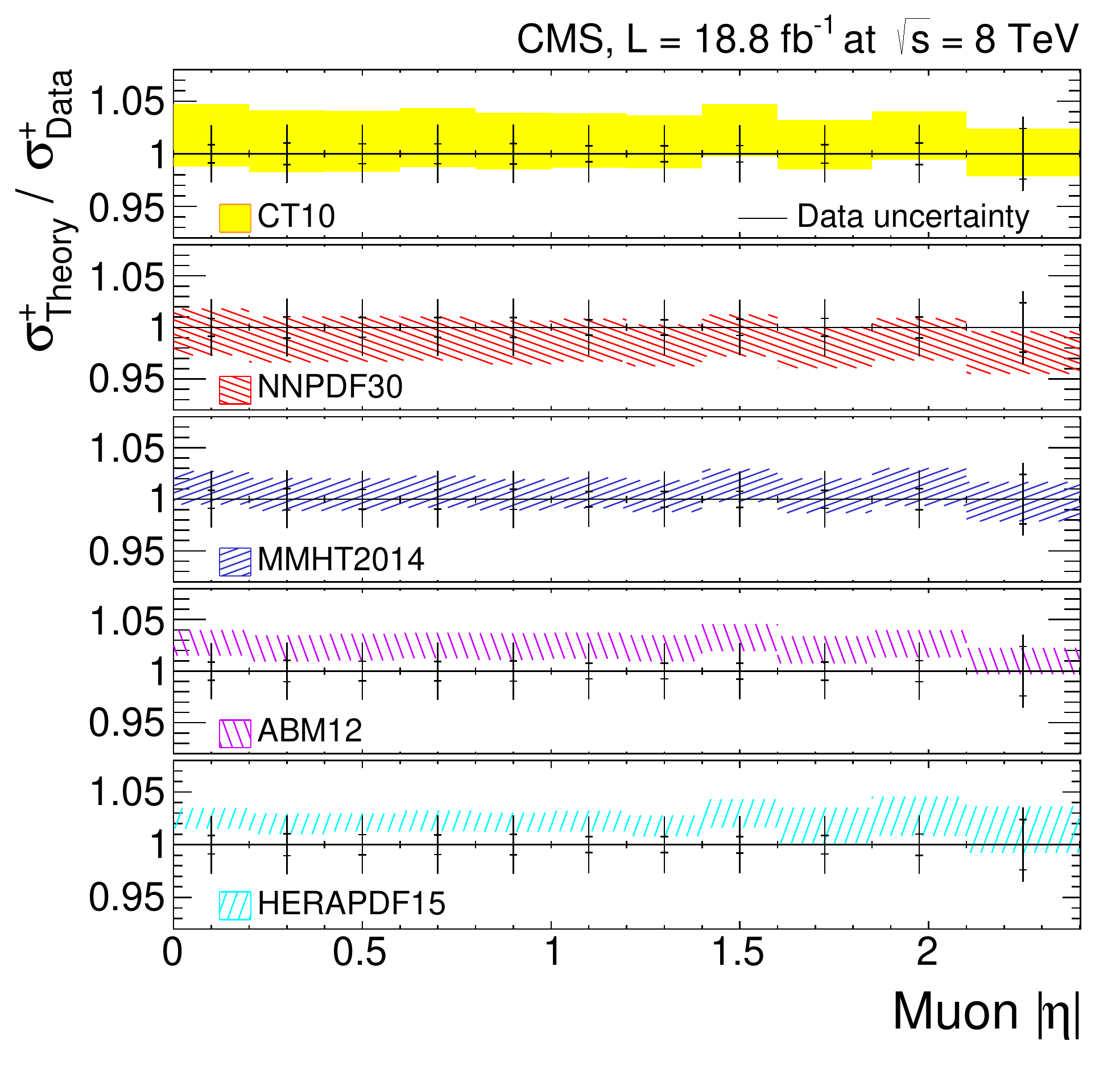}
    \includegraphics[width=\cmsFigWidthNarrow]{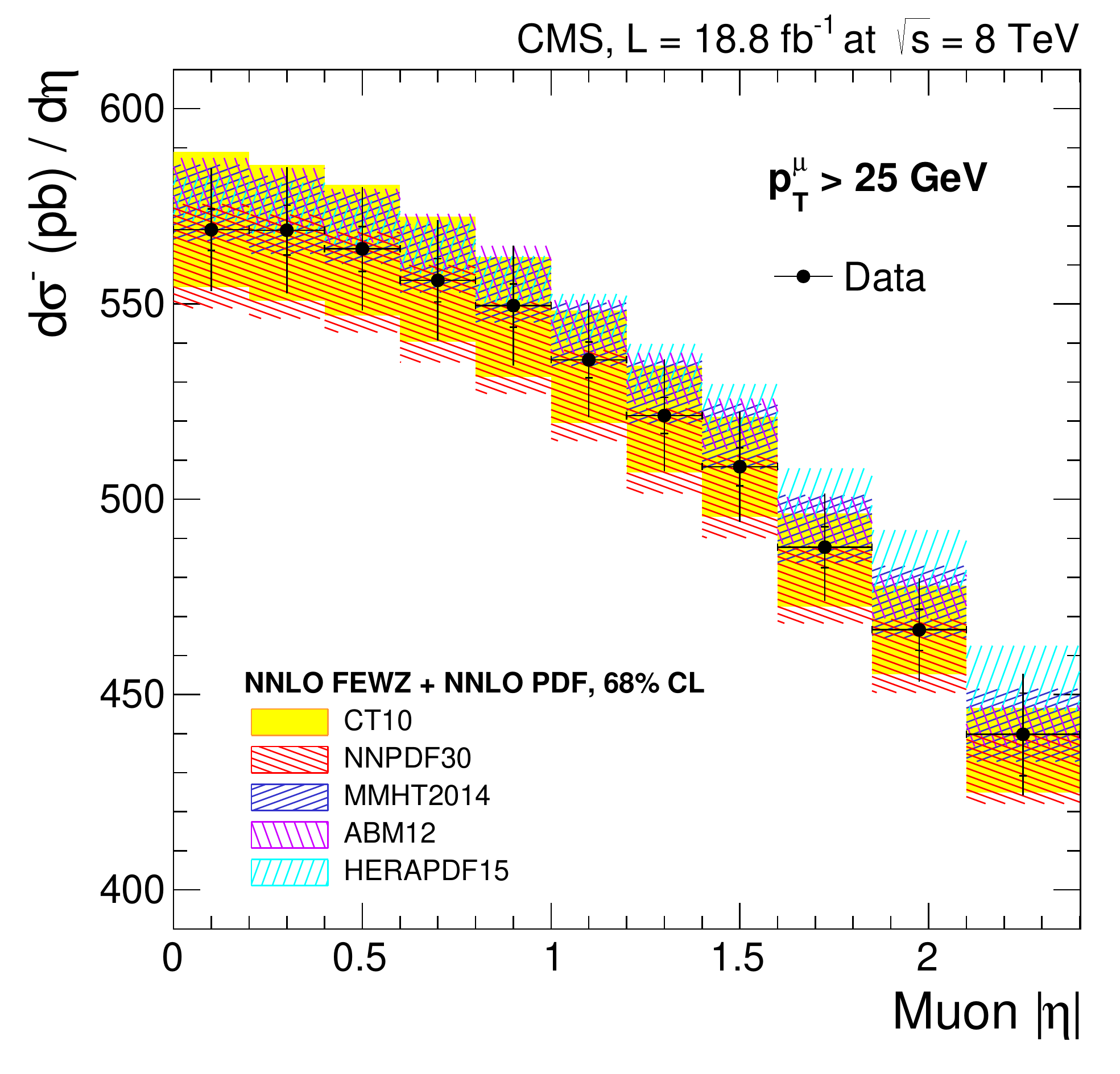}
    \includegraphics[width=\cmsFigWidthNarrow]{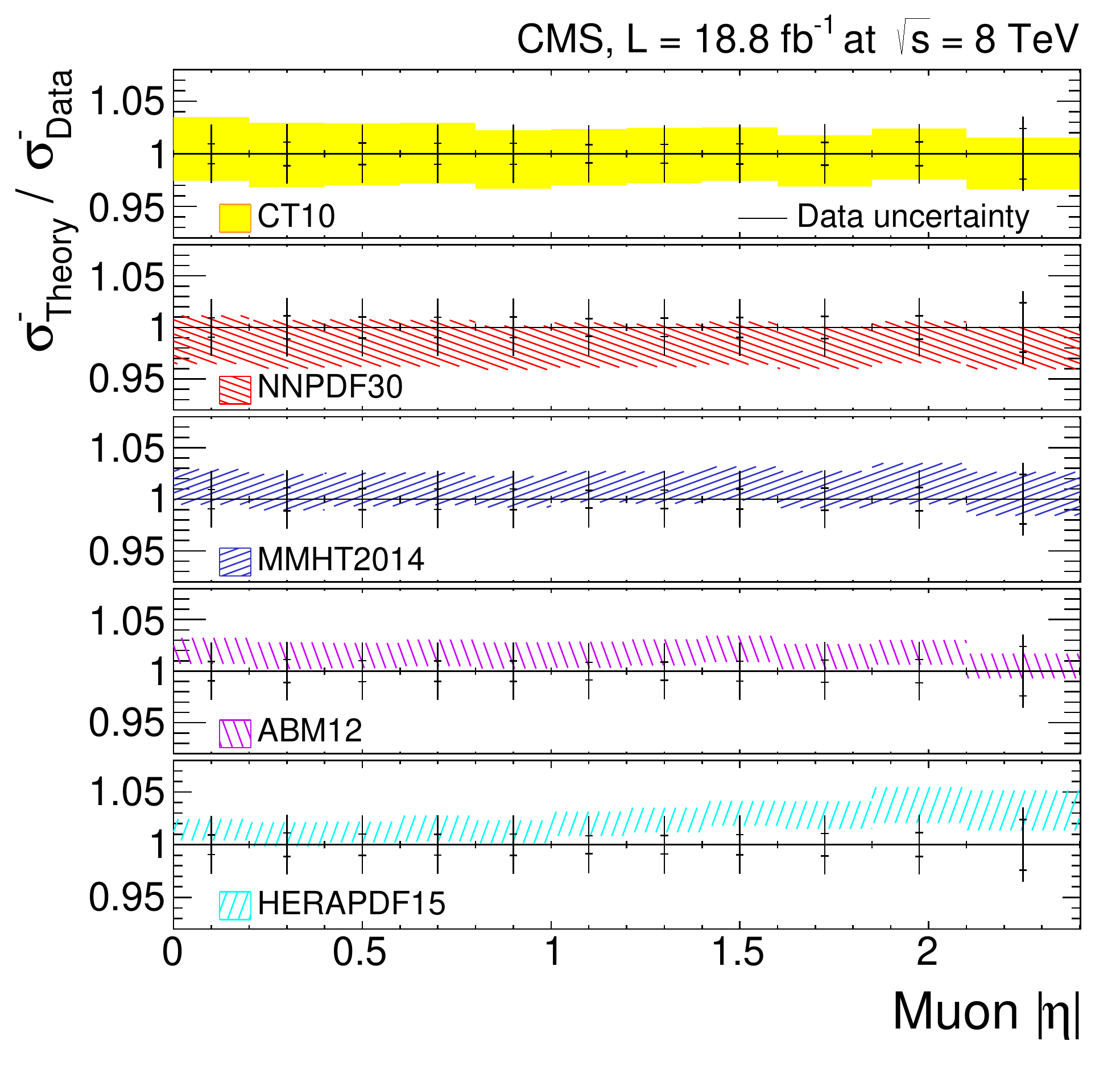}
    \includegraphics[width=\cmsFigWidthNarrow]{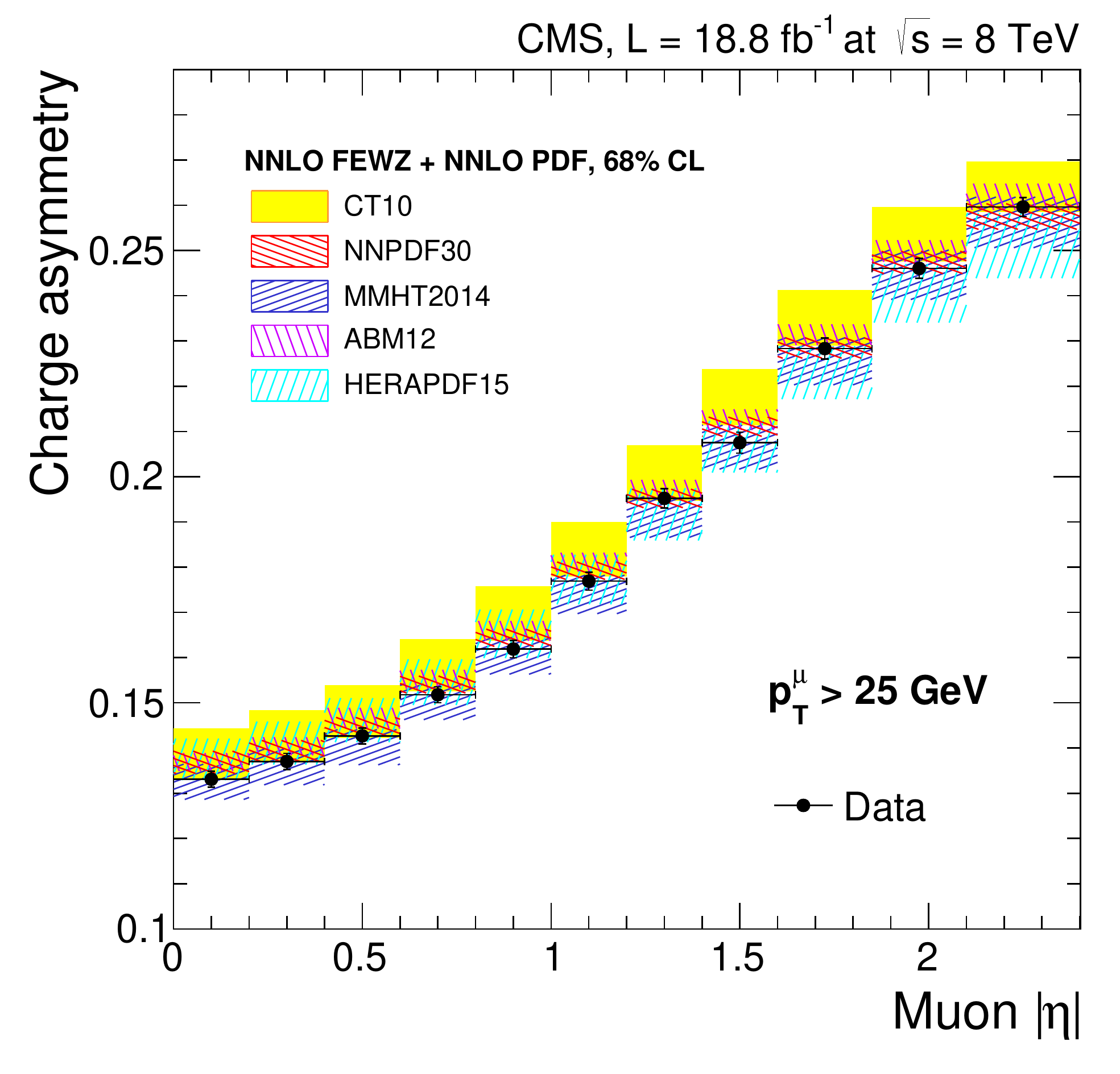}
    \includegraphics[width=\cmsFigWidthNarrow]{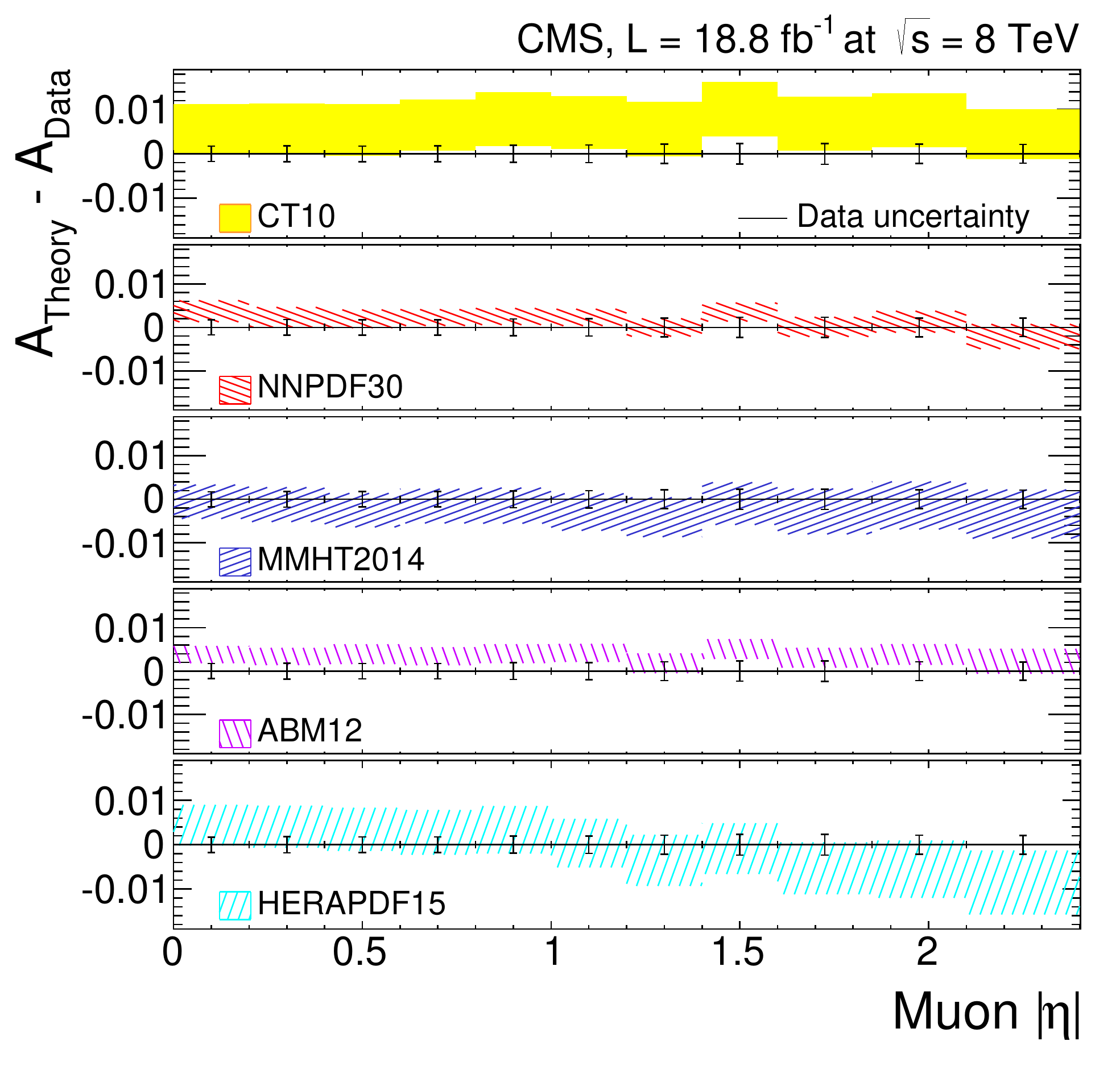}
    \caption{
	Comparison of the measured cross sections (upper plot for $\sigeta^+$ and middle for $\sigeta^-$) and asymmetries (lower plot) to NNLO predictions calculated using the
	\textsc{fewz}~3.1 MC tool interfaced with different PDF sets.
	The right column shows the ratios (differences) between the theoretical predictions and the measured cross sections (asymmetries).
	The smaller vertical error bars on the data points represent the statistical and systematic uncertainties.
	The full error bars include the integrated luminosity uncertainty.
	The PDF uncertainty of each PDF set is shown by a shaded~(or hatched) band and corresponds to 68\% \CL.
	\label{fig:results:final}
    }
\end{figure*}

{\tolerance=5000
The measured cross sections and charge asymmetries are summarized in Table~\ref{table:results:final} and displayed in Fig.~\ref{fig:results:final}.
The error bars of the measurements represent both statistical and systematic uncertainties, including the uncertainty in the integrated luminosity.
The measurements are compared with theoretical predictions based on several PDF sets.
The predictions are obtained using the \textsc{fewz} 3.1 \cite{FEWZ} NNLO MC calculation interfaced with CT10 \cite{CTEQ:1007}, NNPDF3.0 \cite{Ball:2014uwa}, HERA\-PDF1.5 \cite{HERAPDF1_5}, MMHT2014 \cite{MMHT2014}, and ABM12 \cite{Alekhin:2013nda} PDF sets.
No electroweak corrections are included in these calculations.
The error bars of the theoretical predictions represent the PDF uncertainty, which is the dominant source of uncertainty in these calculations.
For the CT10, MMHT, HERA, and ABM PDFs, the uncertainties are calculated with their eigenvector sets using asymmetric master equations where applicable. For the NNPDF set the standard deviations over its 100 replicas are evaluated.

The numerical values of the predictions are also shown in Table~\ref{table:results:final}.
We note that the previous lepton charge asymmetries measured by CMS at $\sqrt{s}=7\TeV$ have been included in the global PDF fits for the NNPDF3.0, MMHT2014, and ABM12 PDFs.
The measured cross sections and charge asymmetries are well described by all considered PDF sets within their corresponding uncertainties.
}

\begin{table*}[htbp]
\centering
\topcaption{
    Summary of the measured differential cross sections $\sigeta^\pm$~(pb) and charge asymmetry~$\mathcal{A}$.
    The first uncertainty is statistical, the second uncertainty is systematic, and the third is the integrated luminosity uncertainty.
    The theoretical predictions are obtained using the \textsc{fewz}~3.1~\cite{FEWZ} NNLO MC tool interfaced with five different PDF sets.
}
\label{table:results:final}
\resizebox{\textwidth}{!}{
\renewcommand\arraystretch{1.25}
\begin{tabular}{ l | c | c  c  c  c  c  }
\hline
		 & Measurement &  \multicolumn{5}{c}{Theory}   \\
$\abs{\eta}$ bin & ($\pm$ stat $\pm$ syst $\pm$ lumi) &  CT10   & NNPDF3.0  &  MMHT2014  & ABM12 &  HERAPDF1.5   \\ \hline
\hline
\multicolumn{7}{c}{$\sigeta^+$ (pb)} \\
\hline
0.00--0.20 & $ 743.7 \pm  0.7 \pm  6.5 \pm 19.3 $ & $ 759.7^{+19.3}_{-25.1} $ & $ 740.5\pm16.8 $ & $ 750.8^{+13.2}_{-10.8} $ & $ 764.2\pm 9.3 $ & $ 762.8^{+ 6.8}_{- 7.8} $  \\
0.20--0.40 & $ 749.5 \pm  0.7 \pm  7.7 \pm 19.5 $ & $ 761.2^{+19.2}_{-24.9} $ & $ 740.8\pm16.6 $ & $ 751.8^{+13.1}_{-10.6} $ & $ 766.0\pm 9.6 $ & $ 764.7^{+ 7.2}_{- 7.8} $  \\
0.40--0.60 & $ 751.9 \pm  0.7 \pm  7.2 \pm 19.5 $ & $ 763.6^{+19.1}_{-24.6} $ & $ 743.5\pm16.5 $ & $ 754.0^{+13.0}_{-10.3} $ & $ 769.4\pm 9.7 $ & $ 767.9^{+ 6.5}_{- 6.6} $  \\
0.60--0.80 & $ 755.0 \pm  0.7 \pm  7.1 \pm 19.6 $ & $ 769.1^{+18.6}_{-23.8} $ & $ 746.9\pm16.0 $ & $ 759.0^{+13.1}_{-10.1} $ & $ 773.8\pm 9.4 $ & $ 772.0^{+ 7.8}_{- 7.2} $  \\
0.80--1.00 & $ 761.9 \pm  0.7 \pm  7.4 \pm 19.8 $ & $ 773.4^{+18.2}_{-22.8} $ & $ 750.7\pm16.0 $ & $ 763.6^{+13.0}_{- 9.8} $ & $ 780.0\pm 9.9 $ & $ 777.5^{+ 7.6}_{- 6.4} $  \\
1.00--1.20 & $ 766.0 \pm  0.7 \pm  5.8 \pm 19.9 $ & $ 777.8^{+17.7}_{-22.1} $ & $ 756.5\pm15.8 $ & $ 769.2^{+12.8}_{- 9.8} $ & $ 784.9\pm 9.7 $ & $ 782.5^{+ 8.2}_{- 6.8} $  \\
1.20--1.40 & $ 774.4 \pm  0.7 \pm  5.8 \pm 20.1 $ & $ 785.0^{+17.7}_{-21.5} $ & $ 760.9\pm15.6 $ & $ 775.5^{+13.1}_{-10.5} $ & $ 791.5\pm 9.9 $ & $ 787.3^{+ 8.7}_{- 6.8} $  \\
1.40--1.60 & $ 774.6 \pm  0.7 \pm  6.1 \pm 20.1 $ & $ 793.7^{+17.5}_{-20.8} $ & $ 768.5\pm15.7 $ & $ 784.0^{+13.3}_{-11.3} $ & $ 799.7\pm10.2 $ & $ 796.7^{+11.4}_{- 9.5} $  \\
1.60--1.85 & $ 776.4 \pm  0.7 \pm  6.8 \pm 20.2 $ & $ 784.4^{+16.9}_{-19.5} $ & $ 761.3\pm15.4 $ & $ 778.5^{+13.5}_{-12.4} $ & $ 792.4\pm10.3 $ & $ 788.9^{+15.0}_{-11.5} $  \\
1.85--2.10 & $ 771.1 \pm  0.6 \pm  7.9 \pm 20.0 $ & $ 785.5^{+16.9}_{-18.8} $ & $ 762.2\pm15.7 $ & $ 780.3^{+14.0}_{-14.0} $ & $ 791.6\pm10.2 $ & $ 788.9^{+17.6}_{-11.4} $  \\
2.10--2.40 & $ 748.3 \pm  0.7 \pm 18.0 \pm 19.5 $ & $ 750.0^{+16.4}_{-17.7} $ & $ 730.1\pm15.4 $ & $ 746.9^{+13.9}_{-14.6} $ & $ 755.6\pm 9.6 $ & $ 754.8^{+20.9}_{-12.3} $  \\
\hline
\hline
\multicolumn{7}{c}{$\sigeta^- $ (pb)} \\
\hline
0.00--0.20 & $ 569.0 \pm  0.6 \pm  5.3 \pm 14.8 $ & $ 574.5^{+14.5}_{-20.2} $ & $ 562.2\pm13.3 $ & $ 576.2^{+ 9.4}_{-10.1} $ & $ 580.2\pm 7.2 $ & $ 578.8^{+ 4.1}_{- 7.6} $  \\
0.20--0.40 & $ 568.9 \pm  0.6 \pm  6.4 \pm 14.8 $ & $ 571.0^{+14.6}_{-20.1} $ & $ 559.6\pm13.3 $ & $ 573.2^{+ 9.6}_{-10.3} $ & $ 577.4\pm 7.4 $ & $ 576.1^{+ 5.0}_{- 8.1} $  \\
0.40--0.60 & $ 564.1 \pm  0.6 \pm  5.7 \pm 14.7 $ & $ 566.4^{+14.2}_{-19.3} $ & $ 555.6\pm12.8 $ & $ 569.7^{+ 8.8}_{- 9.3} $ & $ 572.6\pm 6.9 $ & $ 572.5^{+ 4.1}_{- 7.2} $  \\
0.60--0.80 & $ 556.1 \pm  0.6 \pm  5.5 \pm 14.5 $ & $ 558.6^{+13.7}_{-18.3} $ & $ 547.5\pm12.4 $ & $ 561.8^{+ 8.6}_{- 8.9} $ & $ 565.9\pm 7.2 $ & $ 565.7^{+ 5.8}_{- 8.0} $  \\
0.80--1.00 & $ 549.6 \pm  0.6 \pm  5.5 \pm 14.3 $ & $ 548.6^{+13.4}_{-17.3} $ & $ 538.8\pm11.7 $ & $ 553.6^{+ 8.3}_{- 8.3} $ & $ 557.9\pm 7.0 $ & $ 557.4^{+ 4.9}_{- 7.0} $  \\
1.00--1.20 & $ 535.7 \pm  0.6 \pm  4.6 \pm 13.9 $ & $ 535.6^{+12.8}_{-16.0} $ & $ 526.6\pm11.6 $ & $ 542.2^{+ 8.0}_{- 8.1} $ & $ 544.2\pm 6.8 $ & $ 547.2^{+ 5.3}_{- 7.0} $  \\
1.20--1.40 & $ 521.4 \pm  0.6 \pm  4.6 \pm 13.6 $ & $ 521.8^{+12.4}_{-14.9} $ & $ 512.4\pm10.9 $ & $ 527.5^{+ 8.0}_{- 8.2} $ & $ 530.9\pm 6.6 $ & $ 534.5^{+ 5.3}_{- 7.0} $  \\
1.40--1.60 & $ 508.3 \pm  0.6 \pm  4.9 \pm 13.2 $ & $ 509.3^{+11.8}_{-13.9} $ & $ 500.6\pm10.5 $ & $ 516.3^{+ 8.2}_{- 8.4} $ & $ 519.3\pm 6.5 $ & $ 524.2^{+ 5.4}_{- 6.7} $  \\
1.60--1.85 & $ 487.7 \pm  0.6 \pm  5.2 \pm 12.7 $ & $ 485.1^{+11.2}_{-12.6} $ & $ 478.1\pm 9.9 $ & $ 492.5^{+ 8.6}_{- 8.8} $ & $ 494.6\pm 6.0 $ & $ 501.6^{+ 6.3}_{- 6.6} $  \\
1.85--2.10 & $ 466.6 \pm  0.6 \pm  5.3 \pm 12.1 $ & $ 467.0^{+11.0}_{-11.7} $ & $ 459.9\pm 9.4 $ & $ 473.8^{+ 9.1}_{- 9.4} $ & $ 475.1\pm 5.6 $ & $ 483.4^{+ 8.7}_{- 7.3} $  \\
2.10--2.40 & $ 439.8 \pm  0.6 \pm 10.6 \pm 11.4 $ & $ 436.0^{+10.6}_{-11.1} $ & $ 431.0\pm 9.0 $ & $ 442.3^{+ 9.1}_{- 9.4} $ & $ 442.0\pm 5.3 $ & $ 452.4^{+10.1}_{- 6.6} $  \\

\hline
\hline
\multicolumn{7}{c}{$\mathcal{A}$ (\%)} \\
\hline
0.00--0.20 & $  13.31 \pm  0.06 \pm  0.17 $ & $  13.89^{+ 0.55}_{- 0.57} $ & $  13.68\pm 0.25 $ & $  13.16^{+ 0.48}_{- 0.30} $ & $  13.69\pm 0.20 $ & $  13.71^{+ 0.50}_{- 0.43} $  \\
0.20--0.40 & $  13.70 \pm  0.06 \pm  0.18 $ & $  14.28^{+ 0.56}_{- 0.59} $ & $  13.94\pm 0.23 $ & $  13.48^{+ 0.49}_{- 0.30} $ & $  14.04\pm 0.20 $ & $  14.07^{+ 0.51}_{- 0.44} $  \\
0.40--0.60 & $  14.27 \pm  0.06 \pm  0.18 $ & $  14.83^{+ 0.56}_{- 0.60} $ & $  14.47\pm 0.21 $ & $  13.92^{+ 0.48}_{- 0.30} $ & $  14.66\pm 0.23 $ & $  14.58^{+ 0.53}_{- 0.45} $  \\
0.60--0.80 & $  15.18 \pm  0.06 \pm  0.18 $ & $  15.85^{+ 0.55}_{- 0.61} $ & $  15.40\pm 0.19 $ & $  14.93^{+ 0.49}_{- 0.30} $ & $  15.52\pm 0.21 $ & $  15.42^{+ 0.54}_{- 0.47} $  \\
0.80--1.00 & $  16.19 \pm  0.06 \pm  0.19 $ & $  17.01^{+ 0.57}_{- 0.64} $ & $  16.44\pm 0.19 $ & $  15.95^{+ 0.50}_{- 0.31} $ & $  16.59\pm 0.22 $ & $  16.49^{+ 0.58}_{- 0.50} $  \\
1.00--1.20 & $  17.69 \pm  0.07 \pm  0.20 $ & $  18.44^{+ 0.55}_{- 0.65} $ & $  17.92\pm 0.19 $ & $  17.31^{+ 0.51}_{- 0.34} $ & $  18.11\pm 0.21 $ & $  17.69^{+ 0.58}_{- 0.51} $  \\
1.20--1.40 & $  19.52 \pm  0.07 \pm  0.22 $ & $  20.14^{+ 0.56}_{- 0.67} $ & $  19.52\pm 0.20 $ & $  19.03^{+ 0.53}_{- 0.38} $ & $  19.70\pm 0.23 $ & $  19.13^{+ 0.62}_{- 0.54} $  \\
1.40--1.60 & $  20.75 \pm  0.07 \pm  0.23 $ & $  21.82^{+ 0.56}_{- 0.68} $ & $  21.10\pm 0.21 $ & $  20.59^{+ 0.55}_{- 0.42} $ & $  21.26\pm 0.23 $ & $  20.63^{+ 0.60}_{- 0.54} $  \\
1.60--1.85 & $  22.83 \pm  0.06 \pm  0.23 $ & $  23.57^{+ 0.55}_{- 0.68} $ & $  22.84\pm 0.23 $ & $  22.50^{+ 0.57}_{- 0.48} $ & $  23.14\pm 0.23 $ & $  22.26^{+ 0.63}_{- 0.55} $  \\
1.85--2.10 & $  24.61 \pm  0.06 \pm  0.22 $ & $  25.43^{+ 0.54}_{- 0.67} $ & $  24.74\pm 0.25 $ & $  24.44^{+ 0.57}_{- 0.52} $ & $  24.99\pm 0.24 $ & $  24.01^{+ 0.69}_{- 0.60} $  \\
2.10--2.40 & $  25.96 \pm  0.07 \pm  0.21 $ & $  26.47^{+ 0.50}_{- 0.62} $ & $  25.75\pm 0.28 $ & $  25.61^{+ 0.57}_{- 0.55} $ & $  26.19\pm 0.29 $ & $  25.05^{+ 0.78}_{- 0.67} $  \\
\hline
\end{tabular}
}
\end{table*}

\section{QCD analysis}

The muon charge asymmetry measurements at 8\TeV presented here are used in a QCD analysis at NNLO together with
the combined measurements of neutral- and charged-current cross sections of deep inelastic electron(positron)-proton
scattering (DIS) at HERA~\cite{Abramowicz:2015mha}. The correlations of the experimental uncertainties for the muon charge asymmetry and for the inclusive DIS cross sections are taken into account. The theoretical predictions are calculated at NLO by using the {\sc mcfm 6.8} program~\cite{Campbell:1999ah,Campbell:2010ff}, which is interfaced to {\sc applgrid 1.4.56}~\cite{Carli:2010rw}. The NNLO corrections are obtained by using $k$-factors, defined as ratios of the predictions at NNLO to the ones at NLO, both calculated with the \textsc{fewz}~3.1~\cite{FEWZ} program, using NNLO CT10~\cite{CTEQ:1007} PDFs.

Version 1.1.1 of the open-source QCD fit framework for PDF determination {\sc heraf}itter~\cite{Alekhin:2014irh, herafitter} is
used with the partons evolved by using the Dokshitzer-Gribov-Lipatov-Altarelli-Parisi equations~\cite{Gribov:1972ri,Altarelli:1977zs,Curci:1980uw,Furmanski:1980cm,Moch:2004pa,Vogt:2004mw} at NNLO, as implemented in the {\sc qcdnum 17-00/06} program~\cite{Botje:2010ay}.

The Thorne-Roberts~\cite{Thorne:2006qt,Martin:2009ad} general mass variable flavor number
scheme at NNLO is used for the treatment of heavy-quark contributions with heavy-quark masses $m_{\cPqc} = 1.43\GeV$
and $m_{\cPqb} = 4.5\GeV$. The renormalization and factorization scales are set to $Q$, which denotes the four-momentum transfer in case of the DIS data and the mass of the $\PW$ boson in case of the muon charge asymmetry, respectively.

The strong coupling constant is set to $\alpha_s (m_{\cPZ})$ = 0.118.
The $Q^2$ range of HERA data is restricted to $Q^2 \geq Q^2_{\textrm{min}} = 3.5\GeV^2$ to assure the
applicability of perturbative QCD over the kinematic range of the fit.

The procedure for the determination of the PDFs follows the approach used in the analysis in Ref.~\cite{CMS:asym:2012}.
The parameterised PDFs are the gluon distribution, $x\Pg$, the valence quark distributions, $x\cPqu_v$, $x\cPqd_v$, and
the $\cPqu$-type and $\cPqd$-type anti-quark distributions, $x\overline{U}$, $x\overline{D}$. At the initial scale of the
QCD evolution $Q_0^2 = 1.9\GeV^2$, the PDFs are parametrized as:
\begin{align}
x\Pg(x) &= A_{\Pg} x^{B_{\Pg}}\,(1-x)^{C_{\Pg}}\, (1+D_{\Pg} x) ,
\label{eq:g}\\
x\cPqu_v(x) &= A_{\cPqu_v}x^{B_{\cPqu_v}}\,(1-x)^{C_{\cPqu_v}}\,(1+E_{\cPqu_v}x^2) ,
\label{eq:uv}\\
x\cPqd_v(x) &= A_{\cPqd_v}x^{B_{\cPqd_v}}\,(1-x)^{C_{\cPqd_v}},
\label{eq:dv}\\
x\overline{U}(x)&= A_{\overline{U}}x^{B_{\overline{U}}}\, (1-x)^{C_{\overline{U}}}\, (1+E_{\overline{U}}x^2),
\label{eq:Ubar}\\
x\overline{D}(x)&= A_{\overline{D}}x^{B_{\overline{D}}}\, (1-x)^{C_{\overline{D}}},
\label{eq:Dbar}
\end{align}
with the relations $x\overline{U} = x\cPaqu$ and $x\overline{D} = x\cPaqd + x\cPaqs$ assumed.

The normalization parameters $A_{\cPqu_{\mathrm{v}}}$, $A_{\cPqd_\mathrm{v}}$, and $A_{\cPg}$ are determined by the QCD sum
rules, the $B$ parameter is responsible for small-$x$ behavior of the PDFs, and the parameter $C$ describes the shape of
the distribution as $x\,{\to}\,1$. Additional constraints $B_{\overline{\mathrm{U}}} = B_{\overline{\mathrm{D}}}$ and $A_{\overline{\mathrm{U}}} = A_{\overline{\mathrm{D}}}(1 - f_{\cPqs})$ are imposed with $f_{\cPqs}$ being the strangeness fraction, $f_{\cPqs} = \cPaqs/( \cPaqd + \cPaqs)$,
which is fixed to $f_{\cPqs}=0.31\pm0.08$ as in Ref.~\cite{Martin:2009ad}, consistent with the determination of the
strangeness fraction by using the CMS measurements of $\PW$ + charm production~\cite{CMS:asym:2012}.
The $\chi^2$ definition in the QCD analysis follows that of Eq.(32) of~\cite{Abramowicz:2015mha} without the logarithmic term.
The parameters in Eqs.~(\ref{eq:g})-(\ref{eq:Dbar}) were selected by first fitting with all $D$ and $E$ parameters set to zero.
The other parameters were then included in the fit one at a time independently. The improvement of the $\chi^2$ of the fits was monitored
and the procedure was stopped when no further improvement was observed. This led to a 13-parameter fit.

The PDF uncertainties are estimated according to the general approach of {\sc HERAPDF1.0}~\cite{Aaron:2009aa} in which
the experimental, model, and parametrization uncertainties are taken into account. A tolerance criterion
of $\Delta\chi^2 =1$ is adopted for defining the experimental uncertainties that originate from the measurements
included in the analysis.

Model uncertainties arise from the variations in the values assumed for the heavy-quark masses
$m_{\cPqb}$, $m_{\cPqc}$ with $4.25\leq m_{\cPqb}\leq 4.75\GeV$, $1.37\leq m_{\cPqc}\leq 1.49\GeV$, following Ref.~\cite{Abramowicz:2015mha},
and the value of $Q^2_{\text{min}}$ imposed on the HERA data, which is varied in the interval
$2.5 \leq Q^2_{\text{min}}\leq 5.0\GeV^2$. The strangeness fraction $f_{\cPqs}$ is varied by its uncertainty.

The parametrization uncertainty is estimated by extending the functional form of all parton densities with additional
parameters. The uncertainty is constructed as an envelope built from the maximal differences between the PDFs resulting
from all the parametrization variations and the central fit at each $x$ value.
The total PDF uncertainty is obtained by adding experimental, model, and parametrization uncertainties in quadrature.
In the following, the quoted uncertainties correspond to 68\% CL.

\begin{table*}[!htbp]
\centering
\renewcommand{\arraystretch}{1.25}
\topcaption{Partial $\chi^2$ per number of data points, $n_{\text{dp}}$, and the global $\chi^2$ per
degrees of freedom, $n_{\text{dof}}$, as obtained in the QCD analysis of HERA DIS and the CMS muon charge asymmetry data.
For HERA measurements, the energy of the proton beam is listed for each data set, with electron energy being $E_{\Pe}=27.5\GeV$.}
\begin{tabular}[h]{l | c}
\hline
Data sets   & Partial $\chi^2/n_{\text{dp}}$ \\
\hline
HERA1+2 neutral current,  $\Pep \Pp$,  $E_{\Pp}=920\GeV$ &  $440/377$  \\
HERA1+2 neutral current,  $\Pep \Pp$,  $E_{\Pp}=820\GeV$ &  $69/70$ \\
HERA1+2 neutral current,  $\Pep \Pp$,  $E_{\Pp}=575\GeV$ &  $214/254$ \\
HERA1+2 neutral current,  $\Pep \Pp$,  $E_{\Pp}=460\GeV$ &  $210/204$ \\
HERA1+2 neutral current,  $\Pem \Pp$,  $E_{\Pp}=920\GeV$ &  $218/159$ \\
HERA1+2 charged current,  $\Pep \Pp$,  $E_{\Pp}=920\GeV$ &  $46/39 $ \\
HERA1+2 charged current,  $\Pem \Pp$,  $E_{\Pp}=920\GeV$ &  $50/42 $ \\
Correlated $\chi^2$ of HERA1+2 data                                & $141$ \\
\hline
CMS $\PW^\pm$ muon charge asymmetry   ${\cal A}(\eta_\Pgm)$, $\sqrt{s}=8\TeV$ &  $3/11$ \\
\hline
Global $\chi^2/n_{\text{dof}}$                         & $1391/1143$\\
\hline
\end{tabular}
\label{chi2_paper_table}
\end{table*}

\begin{figure*}[!htbp]
\center
   \includegraphics[width=\cmsFigWidth]{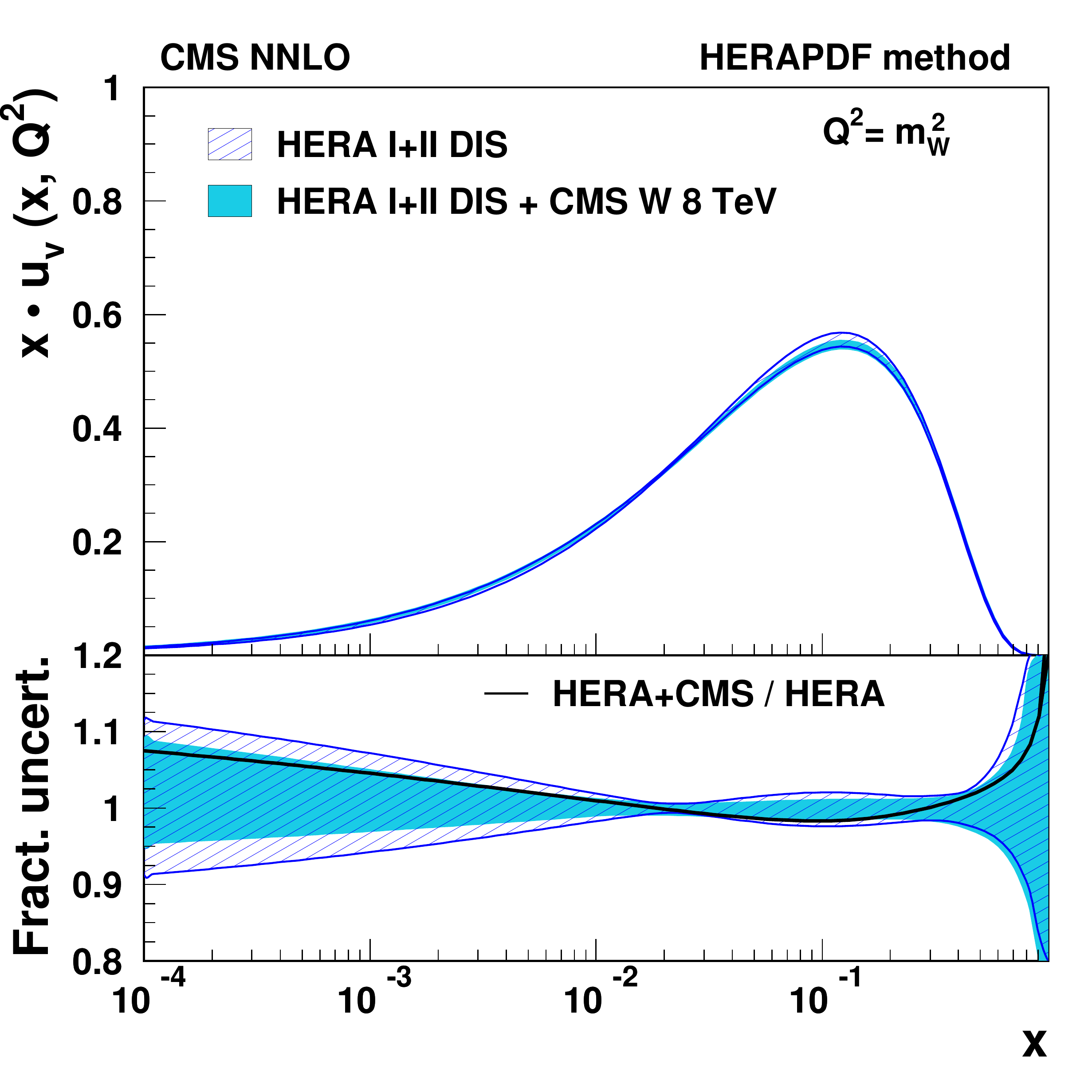}
   \includegraphics[width=\cmsFigWidth]{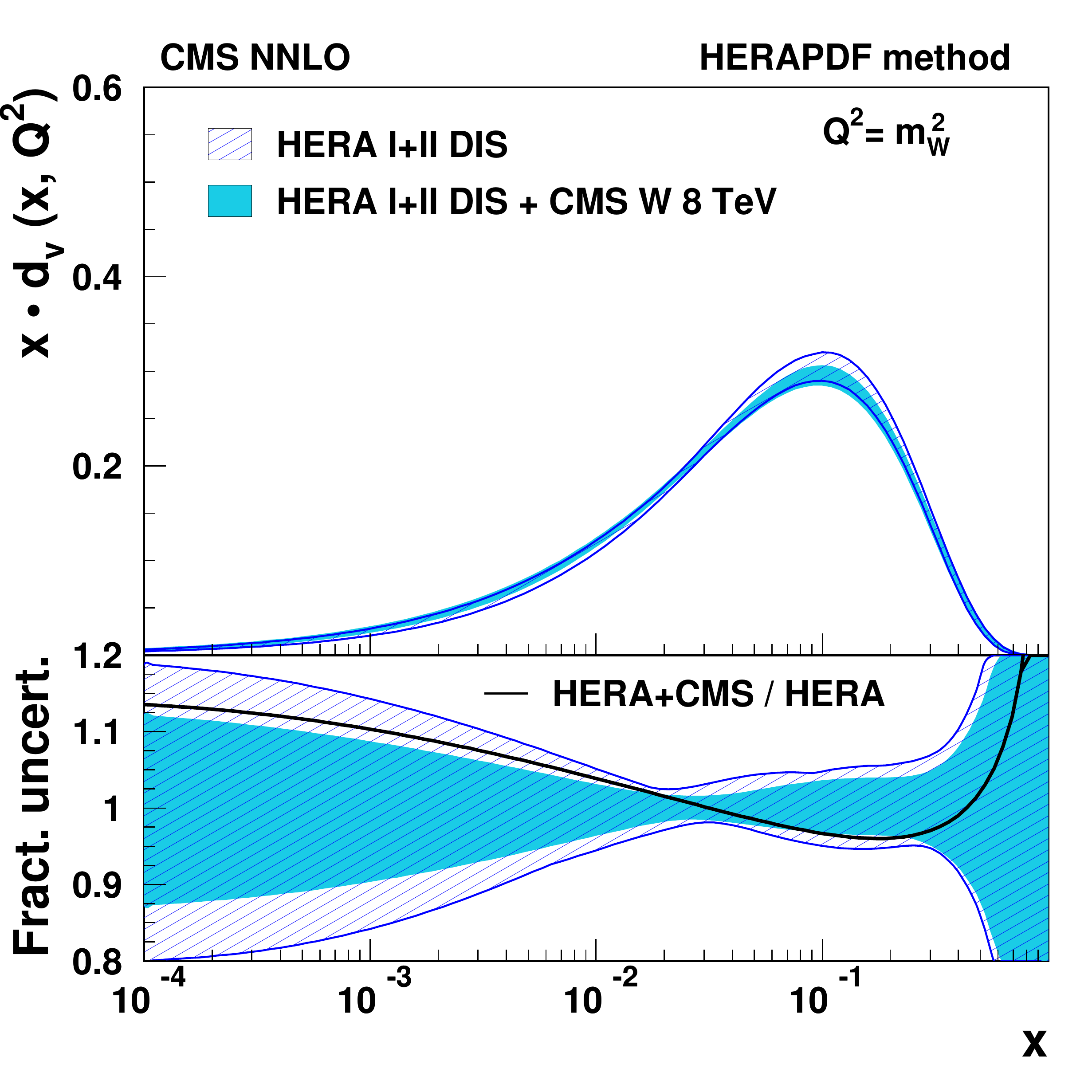}
\caption{
Distributions of $\cPqu$ valence (left) and $\cPqd$ valence (right) quarks as functions of $x$ at the scale $Q^2=m^2_{\PW}$.
The results of the fit to the HERA data and muon asymmetry measurements (light shaded band), and to HERA data only (hatched
band)
are compared. The total PDF uncertainties are shown. In the bottom panels the distributions are normalized to 1 for a direct comparison of the uncertainties.
The change of the PDFs with respect to the HERA-only fit is represented by a solid line.}
\label{herapluswasym}
\end{figure*}

The global and partial $\chi^2$ values for the data sets used are listed in Table~\ref{chi2_paper_table}, illustrating the consistency
among the data sets used. The somewhat high $\chi^2/n_{\mathrm{dof}}$ values for the combined DIS data are very similar to those observed
in Ref.~\cite{Abramowicz:2015mha}, where they are investigated in detail.

{\tolerance=5000
In the kinematic range probed, the final combined HERA DIS data currently provide the most significant constraints on the valence distributions. By adding these muon charge asymmetry measurements, the constraints can be significantly improved, as illustrated in Fig.~\ref{herapluswasym} where the $x \cPqu$ and $x \cPqd$ valence distributions are shown at the scale of $m^2_{\PW}$, relevant for the $\PW$ boson production. The changes in shapes and the reduction of the uncertainties of the valence quark distributions with respect to those obtained with the HERA data are clear.

For direct comparison to the results of the earlier CMS QCD analysis~\cite{CMS:asym:2012} based on the $\PW$ asymmetry measured at $\sqrt{s} = 7\TeV$ and
the subset of HERA DIS data~\cite{Aaron:2009aa}, an alternative PDF fit is performed at NLO, following exactly the data and model inputs
of Ref.~\cite{CMS:asym:2012}, but replacing the CMS measurements at $\sqrt{s} = 7\TeV$ by those at $\sqrt{s} = 8\TeV$. Also, a combined QCD analysis of
both CMS data sets is performed. Very good agreement is observed between the CMS measurements of  $\PW$ asymmetry at $\sqrt{s}=7\TeV$ and $\sqrt{s} = 8\TeV$
and a similar effect on the central values of the PDFs as reported in Ref.~\cite{CMS:asym:2012}. Compared to the PDFs obtained with
HERA only data, the improvement of the precision in the valence quark distributions is more pronounced, when the measurements at $\sqrt{s} = 8\TeV$ are used compared to the results of Ref.~\cite{CMS:asym:2012}. Due to somewhat lower Bjorken $x$ probed by the measurements at $8\TeV$, as compared to $7\TeV$, the
two data sets are complementary and should both be used in the future global QCD analyses.
}

\section{Summary}
In summary, we have measured the differential cross section and charge asymmetry of the $\Wpm \to \PGm^{\pm}\PGn$ production in $\Pp\Pp$ collisions at $\sqrt{s}=8\TeV$ using a data sample corresponding to an integrated luminosity of 18.8\fbinv collected with the CMS detector at the LHC. The measurements were performed in 11 bins of absolute muon pseudorapidity $\etaabs$ for muons with $\pt>25\GeV$. The results have been incorporated into a QCD analysis at next-to-next-to-leading-order together with the inclusive
deep inelastic scattering data from HERA. A significant improvement in the accuracy of the valence quark distributions is observed in the range $10^{-3} < x <10^{-1}$, demonstrating the power of these muon charge asymmetry measurements to improve the main constraints on the valence distributions imposed by the HERA data, in the kinematics range probed. This strongly suggests the use of these measurements in future PDF determinations.

\begin{acknowledgments}

We congratulate our colleagues in the CERN accelerator departments for the excellent performance of the LHC and thank the technical and administrative staffs at CERN and at other CMS institutes for their contributions to the success of the CMS effort. In addition, we gratefully acknowledge the computing centres and personnel of the Worldwide LHC Computing Grid for delivering so effectively the computing infrastructure essential to our analyses. Finally, we acknowledge the enduring support for the construction and operation of the LHC and the CMS detector provided by the following funding agencies: BMWFW and FWF (Austria); FNRS and FWO (Belgium); CNPq, CAPES, FAPERJ, and FAPESP (Brazil); MES (Bulgaria); CERN; CAS, MoST, and NSFC (China); COLCIENCIAS (Colombia); MSES and CSF (Croatia); RPF (Cyprus); MoER, ERC IUT and ERDF (Estonia); Academy of Finland, MEC, and HIP (Finland); CEA and CNRS/IN2P3 (France); BMBF, DFG, and HGF (Germany); GSRT (Greece); OTKA and NIH (Hungary); DAE and DST (India); IPM (Iran); SFI (Ireland); INFN (Italy); MSIP and NRF (Republic of Korea); LAS (Lithuania); MOE and UM (Malaysia); CINVESTAV, CONACYT, SEP, and UASLP-FAI (Mexico); MBIE (New Zealand); PAEC (Pakistan); MSHE and NSC (Poland); FCT (Portugal); JINR (Dubna); MON, RosAtom, RAS and RFBR (Russia); MESTD (Serbia); SEIDI and CPAN (Spain); Swiss Funding Agencies (Switzerland); MST (Taipei); ThEPCenter, IPST, STAR and NSTDA (Thailand); TUBITAK and TAEK (Turkey); NASU and SFFR (Ukraine); STFC (United Kingdom); DOE and NSF (USA).

Individuals have received support from the Marie-Curie programme and the European Research Council and EPLANET (European Union); the Leventis Foundation; the A. P. Sloan Foundation; the Alexander von Humboldt Foundation; the Belgian Federal Science Policy Office; the Fonds pour la Formation \`a la Recherche dans l'Industrie et dans l'Agriculture (FRIA-Belgium); the Agentschap voor Innovatie door Wetenschap en Technologie (IWT-Belgium); the Ministry of Education, Youth and Sports (MEYS) of the Czech Republic; the Council of Science and Industrial Research, India; the HOMING PLUS programme of the Foundation for Polish Science, cofinanced from European Union, Regional Development Fund; the OPUS programme of the National Science Center (Poland); the Compagnia di San Paolo (Torino); MIUR project 20108T4XTM (Italy); the Thalis and Aristeia programmes cofinanced by EU-ESF and the Greek NSRF; the National Priorities Research Program by Qatar National Research Fund; the Rachadapisek Sompot Fund for Postdoctoral Fellowship, Chulalongkorn University (Thailand); the Chulalongkorn Academic into Its 2nd Century Project Advancement Project (Thailand); and the Welch Foundation, contract C-1845.

\end{acknowledgments}

\bibliography{auto_generated}

\providecommand{\href}[2]{#2}\begingroup\raggedright\begin{thebibliography}{10}%
\makeatletter
\providecommand{\hrefCMSnoop }[0]{\@secondoftwo}%
\makeatother
\providecommand{\doi}{\texttt{doi:}\begingroup \urlstyle{tt}\Url}

\bibitem{Bjorken}
\hrefCMSnoop {}{J.~D. Bjorken and E.~A. Paschos, ``{Inelastic electron--proton
  and $\gamma$--proton scattering and the structure of the nucleon}'',}
  \textit{ Phys. Rev.} \textbf{ 185} (1969) 1975,
\href{http://dx.doi.org/10.1103/PhysRev.185.1975}{\doi{10.1103/PhysRev.185.1975}}.

\bibitem{CDF:wasym1}
\hrefCMSnoop {}{{CDF} Collaboration, ``{Measurement of the lepton charge
  asymmetry in $\PW$ boson decays produced in $\Pp \PAp$ collisions}'',}
  \textit{ Phys. Rev. Lett.} \textbf{ 81} (1998) 5754,
  \href{http://dx.doi.org/10.1103/PhysRevLett.81.5754}{\doi{10.1103/PhysRevLett.81.5754}},
\href{http://www.arXiv.org/abs/hep-ex/9809001}{\texttt{arXiv:hep-ex/9809001}}.

\bibitem{CDF:wasym2}
\hrefCMSnoop {}{{CDF} Collaboration, ``{Direct measurement of the $\PW$
  production charge asymmetry in $ \Pp \PAp $ Collisions at $\sqrt{s} =
  1.96\TeV$}'',} \textit{ Phys. Rev. Lett.} \textbf{ 102} (2009) 181801,
  \href{http://dx.doi.org/10.1103/PhysRevLett.102.181801}{\doi{10.1103/PhysRevLett.102.181801}},
\href{http://www.arXiv.org/abs/0901.2169}{\texttt{arXiv:0901.2169}}.

\bibitem{Abazov:2007pm}
\hrefCMSnoop {}{{D0} Collaboration, ``{Measurement of the muon charge asymmetry
  from $\PW$ boson decays}'',} \textit{ Phys. Rev. D} \textbf{ 77} (2008)
  011106,
  \href{http://dx.doi.org/10.1103/PhysRevD.77.011106}{\doi{10.1103/PhysRevD.77.011106}},
\href{http://www.arXiv.org/abs/0709.4254}{\texttt{arXiv:0709.4254}}.

\bibitem{Abazov:2008qv}
\hrefCMSnoop {}{{D0} Collaboration, ``{Measurement of the electron charge
  asymmetry in $\Pp \PAp \to \PW + X \to \Pe \nu + X $ events at $\sqrt{s} =
  1.96\TeV$}'',} \textit{ Phys. Rev. Lett.} \textbf{ 101} (2008) 211801,
  \href{http://dx.doi.org/10.1103/PhysRevLett.101.211801}{\doi{10.1103/PhysRevLett.101.211801}},
\href{http://www.arXiv.org/abs/0807.3367}{\texttt{arXiv:0807.3367}}.

\bibitem{Abazov:2013rja}
\hrefCMSnoop {}{{D0} Collaboration, ``{Measurement of the muon charge asymmetry
  in $\Pp \PAp \to \PW + X \to \mu \nu + X $ events at $\sqrt{s} =
  1.96\TeV$}'',} \textit{ Phys. Rev. D} \textbf{ 88} (2013) 091102,
  \href{http://dx.doi.org/10.1103/PhysRevD.88.091102}{\doi{10.1103/PhysRevD.88.091102}},
\href{http://www.arXiv.org/abs/1309.2591}{\texttt{arXiv:1309.2591}}.

\bibitem{CMS:asym:2010}
\hrefCMSnoop {}{{{CMS}} Collaboration, ``Measurement of the lepton charge
  asymmetry in inclusive $\PW$ production in $\Pp\Pp$ collisions at $\sqrt{s} =
  7\TeV$'',} \textit{ JHEP} \textbf{ 04} (2011) 050,
  \href{http://dx.doi.org/10.1007/JHEP04(2011)050}{\doi{10.1007/JHEP04(2011)050}},
  \href{http://www.arXiv.org/abs/1103.3470}{\texttt{arXiv:1103.3470}}.

\bibitem{PhysRevD.85.072004}
\hrefCMSnoop {}{{ATLAS Collaboration}, ``{Measurement of the inclusive
  $\PW^\pm$ and $\PZ/\gamma$ cross sections in the $\Pe$ and $\mu$ decay
  channels in $\Pp\Pp$ collisions at $\sqrt{s} = 7\TeV$ with the ATLAS
  detector}'',} \textit{ Phys. Rev. D} \textbf{ 85} (2012) 072004,
  \href{http://dx.doi.org/10.1103/PhysRevD.85.072004}{\doi{10.1103/PhysRevD.85.072004}},
  \href{http://www.arXiv.org/abs/1109.5141}{\texttt{arXiv:1109.5141}}.

\bibitem{LHCb:wz:2010}
\hrefCMSnoop {}{{{LHCb}} Collaboration, ``Inclusive $\PW$ and $\PZ$ production
  in the forward region at $\sqrt{s} = 7\TeV$'',} \textit{ JHEP} \textbf{ 06}
  (2012) 058,
  \href{http://dx.doi.org/10.1007/JHEP06(2012)058}{\doi{10.1007/JHEP06(2012)058}},
  \href{http://www.arXiv.org/abs/1204.1620}{\texttt{arXiv:1204.1620}}.

\bibitem{CMS:asym:2011}
\hrefCMSnoop {}{{CMS Collaboration}, ``{Measurement of the Electron Charge
  Asymmetry in Inclusive $\PW$ Production in $\Pp\Pp$ Collisions at $\sqrt{s} =
  7\TeV$}'',} \textit{ Phys. Rev. Lett.} \textbf{ 109} (2012) 111806,
  \href{http://dx.doi.org/10.1103/PhysRevLett.109.111806}{\doi{10.1103/PhysRevLett.109.111806}},
\href{http://www.arXiv.org/abs/1206.2598}{\texttt{arXiv:1206.2598}}.

\bibitem{CMS:asym:2012}
\hrefCMSnoop {}{{CMS Collaboration}, ``{Measurement of the muon charge
  asymmetry in inclusive $\Pp\Pp \rightarrow \PW+X$ production at $\sqrt s =
  7\TeV$ and an improved determination of light parton distribution
  functions}'',} \textit{ Phys. Rev. D} \textbf{ 90} (2014) 032004,
  \href{http://dx.doi.org/10.1103/PhysRevD.90.032004}{\doi{10.1103/PhysRevD.90.032004}},
\href{http://www.arXiv.org/abs/1312.6283}{\texttt{arXiv:1312.6283}}.

\bibitem{Chatrchyan:2008zzk}
\hrefCMSnoop {}{{CMS Collaboration}, ``The {CMS} experiment at the {CERN}
  {LHC}'',} \textit{ JINST} \textbf{ 3} (2008) S08004,
\href{http://dx.doi.org/10.1088/1748-0221/3/08/S08004}{\doi{10.1088/1748-0221/3/08/S08004}}.

\bibitem{POWHEG0}
\hrefCMSnoop {}{S.~Alioli, P.~Nason, C.~Oleari, and E.~Re, ``{NLO vector-boson
  production matched with shower in POWHEG}'',} \textit{ JHEP} \textbf{ 07}
  (2008) 060,
  \href{http://dx.doi.org/10.1088/1126-6708/2008/07/060}{\doi{10.1088/1126-6708/2008/07/060}},
\href{http://www.arXiv.org/abs/0805.4802}{\texttt{arXiv:0805.4802}}.

\bibitem{POWHEG1}
\hrefCMSnoop {}{P.~Nason, ``{A new method for combining NLO QCD with shower
  Monte Carlo algorithms}'',} \textit{ JHEP} \textbf{ 11} (2004) 040,
  \href{http://dx.doi.org/10.1088/1126-6708/2004/11/040}{\doi{10.1088/1126-6708/2004/11/040}},
\href{http://www.arXiv.org/abs/hep-ph/0409146}{\texttt{arXiv:hep-ph/0409146}}.

\bibitem{POWHEG2}
\hrefCMSnoop {}{S.~Frixione, P.~Nason, and C.~Oleari, ``{Matching NLO QCD
  computations with parton shower simulations: the POWHEG method}'',} \textit{
  JHEP} \textbf{ 11} (2007) 070,
  \href{http://dx.doi.org/10.1088/1126-6708/2007/11/070}{\doi{10.1088/1126-6708/2007/11/070}},
\href{http://www.arXiv.org/abs/0709.2092}{\texttt{arXiv:0709.2092}}.

\bibitem{POWHEG3}
\hrefCMSnoop {}{S.~Alioli, P.~Nason, C.~Oleari, and E.~Re, ``{A general
  framework for implementing NLO calculations in shower Monte Carlo programs:
  the POWHEG BOX}'',} \textit{ JHEP} \textbf{ 06} (2010) 043,
  \href{http://dx.doi.org/10.1007/JHEP06(2010)043}{\doi{10.1007/JHEP06(2010)043}},
\href{http://www.arXiv.org/abs/1002.2581}{\texttt{arXiv:1002.2581}}.

\bibitem{PYTHIA6}
\hrefCMSnoop {}{T.~Sj{\"o}strand, S.~Mrenna, and P.~Skands, ``{PYTHIA} 6.4
  physics and manual'',} \textit{ JHEP} \textbf{ 05} (2006) 026,
  \href{http://dx.doi.org/10.1088/1126-6708/2006/05/026}{\doi{10.1088/1126-6708/2006/05/026}},
\href{http://www.arXiv.org/abs/hep-ph/0603175}{\texttt{arXiv:hep-ph/0603175}}.

\bibitem{CTEQ:1007}
H.-L. Lai\hrefCMSnoop {}{ {et~al.}, ``{New parton distributions for collider
  physics}'',} \textit{ Phys. Rev. D} \textbf{ 82} (2010) 074024,
  \href{http://dx.doi.org/10.1103/PhysRevD.82.074024}{\doi{10.1103/PhysRevD.82.074024}},
\href{http://www.arXiv.org/abs/1007.2241}{\texttt{arXiv:1007.2241}}.

\bibitem{TAUOLA}
N.~Davidson\hrefCMSnoop {}{ {et~al.}, ``{Universal Interface of TAUOLA
  Technical and Physics Documentation}'',} \textit{ Comput. Phys. Commun.}
  \textbf{ 183} (2012) 821,
  \href{http://dx.doi.org/10.1016/j.cpc.2011.12.009}{\doi{10.1016/j.cpc.2011.12.009}},
\href{http://www.arXiv.org/abs/1002.0543}{\texttt{arXiv:1002.0543}}.

\bibitem{CTEQ6L}
J.~Pumplin\hrefCMSnoop {}{ {et~al.}, ``{New generation of parton distributions
  with uncertainties from global QCD analysis}'',} \textit{ JHEP} \textbf{ 07}
  (2002) 012,
  \href{http://dx.doi.org/10.1088/1126-6708/2002/07/012}{\doi{10.1088/1126-6708/2002/07/012}},
\href{http://www.arXiv.org/abs/hep-ph/0201195}{\texttt{arXiv:hep-ph/0201195}}.

\bibitem{GEANT4}
\hrefCMSnoop {}{{GEANT4} Collaboration, ``{GEANT4---a simulation toolkit}'',}
  \textit{ Nucl. Instrum. Meth. A} \textbf{ 506} (2003) 250,
\href{http://dx.doi.org/10.1016/S0168-9002(03)01368-8}{\doi{10.1016/S0168-9002(03)01368-8}}.

\bibitem{bib:momcor}
A.~Bodek\hrefCMSnoop {}{ {et~al.}, ``{Extracting muon momentum scale
  corrections for hadron collider experiments}'',} \textit{ Eur. Phys. J. C}
  \textbf{ 72} (2012) 2194,
  \href{http://dx.doi.org/10.1140/epjc/s10052-012-2194-8}{\doi{10.1140/epjc/s10052-012-2194-8}},
\href{http://www.arXiv.org/abs/1208.3710}{\texttt{arXiv:1208.3710}}.

\bibitem{CMS:WZ}
\hrefCMSnoop {}{{CMS Collaboration}, ``{Measurements of inclusive $\PW$ and
  $\PZ$ cross sections in $\Pp\Pp$ collisions at $\sqrt{s}=7\TeV$ with the CMS
  experiment}'',} \textit{ JHEP} \textbf{ 10} (2011) 132,
  \href{http://dx.doi.org/10.1007/JHEP10(2011)132}{\doi{10.1007/JHEP10(2011)132}},
\href{http://www.arXiv.org/abs/1107.4789}{\texttt{arXiv:1107.4789}}.

\bibitem{CMS-PAPERS-MUO-10-004}
\hrefCMSnoop {}{{CMS Collaboration}, ``{Performance of CMS muon reconstruction
  in $\Pp\Pp$ collision events at $\sqrt{s}=7\TeV$}'',} \textit{ JINST}
  \textbf{ 7} (2012) P10002,
  \href{http://dx.doi.org/10.1088/1748-0221/7/10/P10002}{\doi{10.1088/1748-0221/7/10/P10002}},
\href{http://www.arXiv.org/abs/1206.4071}{\texttt{arXiv:1206.4071}}.

\bibitem{pflow1}
\href {https://cds.cern.ch/record/1194487}{{CMS Collaboration},
  ``{Particle-flow event reconstruction in CMS and performance for jets, taus,
  and \MET}'',} CMS Physics Analysis Summary CMS-PAS-PFT-09-001, CERN, 2009.

\bibitem{pflow2}
\href {https://cds.cern.ch/record/1247373}{{CMS Collaboration},
  ``{Commissioning of the particle-flow event reconstruction with the first LHC
  collisions recorded in the CMS detector}'',} CMS Physics Analysis Summary
  CMS-PAS-PFT-10-001, CERN, 2010.

\bibitem{t0met}
\hrefCMSnoop {}{{CMS Collaboration}, ``{Performance of the CMS missing
  transverse momentum reconstruction in pp data at $\sqrt{s} = 8\TeV$}'',}
  \textit{ JINST} \textbf{ 10} (2015) P02006,
  \href{http://dx.doi.org/10.1088/1748-0221/10/02/P02006}{\doi{10.1088/1748-0221/10/02/P02006}},
\href{http://www.arXiv.org/abs/1411.0511}{\texttt{arXiv:1411.0511}}.

\bibitem{Abazov:2009tra}
\hrefCMSnoop {}{{D0} Collaboration, ``{A novel method for modeling the recoil
  in $\PW$ boson events at hadron collider}'',} \textit{ Nucl. Instrum. Meth.
  A} \textbf{ 609} (2009) 250,
  \href{http://dx.doi.org/10.1016/j.nima.2009.08.056}{\doi{10.1016/j.nima.2009.08.056}},
\href{http://www.arXiv.org/abs/0907.3713}{\texttt{arXiv:0907.3713}}.

\bibitem{cmsmet}
\hrefCMSnoop {}{{CMS Collaboration}, ``{Missing transverse energy performance
  of the CMS detector}'',} \textit{ JINST} \textbf{ 6} (2011) P09001,
  \href{http://dx.doi.org/10.1088/1748-0221/6/09/P09001}{\doi{10.1088/1748-0221/6/09/P09001}},
\href{http://www.arXiv.org/abs/1106.5048}{\texttt{arXiv:1106.5048}}.

\bibitem{toppp}
\hrefCMSnoop {}{M.~Czakon and A.~Mitov, ``{Top++: a program for the calculation
  of the top-pair cross-section at hadron colliders}'',} \textit{ Comput. Phys.
  Commun.} \textbf{ 185} (2014) 2930,
  \href{http://dx.doi.org/10.1016/j.cpc.2014.06.021}{\doi{10.1016/j.cpc.2014.06.021}},
\href{http://www.arXiv.org/abs/1112.5675}{\texttt{arXiv:1112.5675}}.

\bibitem{Agashe:2014kda}
\hrefCMSnoop {}{{Particle Data Group} Collaboration, ``{Review of particle
  physics}'',} \textit{ Chin. Phys. C} \textbf{ 38} (2014) 090001,
\href{http://dx.doi.org/10.1088/1674-1137/38/9/090001}{\doi{10.1088/1674-1137/38/9/090001}}.

\bibitem{photos}
\hrefCMSnoop {}{G.~Nanava and Z.~Was, ``{How to use SANC to improve the PHOTOS
  Monte Carlo simulation of bremsstrahlung in leptonic $\PW$ boson decays}'',}
  \textit{ Acta Phys. Polon. B} \textbf{ 34} (2003) 4561,
\href{http://www.arXiv.org/abs/hep-ph/0303260}{\texttt{arXiv:hep-ph/0303260}}.

\bibitem{Burkhardt:2001xp}
\hrefCMSnoop {}{H.~Burkhardt and B.~Pietrzyk, ``{Update of the hadronic
  contribution to the QED vacuum polarization}'',} \textit{ Phys. Lett. B}
  \textbf{ 513} (2001) 46,
\href{http://dx.doi.org/10.1016/S0370-2693(01)00393-8}{\doi{10.1016/S0370-2693(01)00393-8}}.

\bibitem{Martin:2009ad}
\hrefCMSnoop {}{A.~D. Martin, W.~J. Stirling, R.~S. Thorne, and G.~Watt,
  ``{Parton distributions for the LHC}'',} \textit{ Eur. Phys. J. C} \textbf{
  63} (2009) 189,
  \href{http://dx.doi.org/10.1140/epjc/s10052-009-1072-5}{\doi{10.1140/epjc/s10052-009-1072-5}},
\href{http://www.arXiv.org/abs/0901.0002}{\texttt{arXiv:0901.0002}}.

\bibitem{Ball:2011mu}
\hrefCMSnoop {}{{NNPDF} Collaboration, ``{Impact of heavy quark masses on
  parton distributions and LHC phenomenology}'',} \textit{ Nucl. Phys. B}
  \textbf{ 849} (2011) 296,
  \href{http://dx.doi.org/10.1016/j.nuclphysb.2011.03.021}{\doi{10.1016/j.nuclphysb.2011.03.021}},
\href{http://www.arXiv.org/abs/1101.1300}{\texttt{arXiv:1101.1300}}.

\bibitem{CMS-PAS-LUM-13-001}
\href {https://cds.cern.ch/record/1598864}{{CMS Collaboration}, ``{CMS
  luminosity based on pixel cluster counting -- Summer 2013 update}'',} CMS
  Physics Analysis Summary CMS-PAS-LUM-13-001, CERN, 2013.

\bibitem{FEWZ}
\hrefCMSnoop {}{Y.~Li and F.~Petriello, ``{Combining QCD and electroweak
  corrections to dilepton production in the framework of the FEWZ simulation
  code}'',} \textit{ Phys. Rev. D} \textbf{ 86} (2012) 094034,
  \href{http://dx.doi.org/10.1103/PhysRevD.86.094034}{\doi{10.1103/PhysRevD.86.094034}},
\href{http://www.arXiv.org/abs/1208.5967}{\texttt{arXiv:1208.5967}}.

\bibitem{Ball:2014uwa}
\hrefCMSnoop {}{{NNPDF} Collaboration, ``{Parton distributions for the LHC Run
  II}'',} \textit{ JHEP} \textbf{ 04} (2015) 040,
  \href{http://dx.doi.org/10.1007/JHEP04(2015)040}{\doi{10.1007/JHEP04(2015)040}},
\href{http://www.arXiv.org/abs/1410.8849}{\texttt{arXiv:1410.8849}}.

\bibitem{HERAPDF1_5}
\href
  {http://pos.sissa.it/archive/conferences/120/168/ICHEP%202010_168.pdf}{{H1
  and ZEUS Collaborations}, ``{Combination and QCD Analysis of the HERA
  Inclusive Cross Sections}'',} in \textit{ Proc. of the 35th Int. Conf. of
  High Energy Physics}, volume 168.
\newblock 2010.
\newblock {PoS}(ICHEP2010)168. Grids available at
  \url{http://www.desy.de/h1zeus/combined_results/index.php?do=proton_structure}.
  Accessed July 2010.

\bibitem{MMHT2014}
\hrefCMSnoop {}{L.~A. Harland-Lang, A.~D. Martin, P.~Motylinski, and R.~S.
  Thorne, ``{Parton distributions in the LHC era: MMHT 2014 PDFs}'',} \textit{
  Eur. Phys. J. C} \textbf{ 75} (2015) 204,
  \href{http://dx.doi.org/10.1140/epjc/s10052-015-3397-6}{\doi{10.1140/epjc/s10052-015-3397-6}},
\href{http://www.arXiv.org/abs/1412.3989}{\texttt{arXiv:1412.3989}}.

\bibitem{Alekhin:2013nda}
\hrefCMSnoop {}{S.~Alekhin, J.~Blumlein, and S.~Moch, ``{The ABM parton
  distributions tuned to LHC data}'',} \textit{ Phys. Rev. D} \textbf{ 89}
  (2014) 054028,
  \href{http://dx.doi.org/10.1103/PhysRevD.89.054028}{\doi{10.1103/PhysRevD.89.054028}},
\href{http://www.arXiv.org/abs/1310.3059}{\texttt{arXiv:1310.3059}}.

\bibitem{Abramowicz:2015mha}
\hrefCMSnoop {}{{H1 and ZEUS Collaborations}, ``{Combination of measurements of
  inclusive deep inelastic ${\Pe^{\pm }\Pp}$ scattering cross sections and QCD
  analysis of HERA data}'',} \textit{ Eur. Phys. J. C} \textbf{ 75} (2015) 580,
  \href{http://dx.doi.org/10.1140/epjc/s10052-015-3710-4}{\doi{10.1140/epjc/s10052-015-3710-4}},
\href{http://www.arXiv.org/abs/1506.06042}{\texttt{arXiv:1506.06042}}.

\bibitem{Campbell:1999ah}
\hrefCMSnoop {}{J.~M. Campbell and R.~K. Ellis, ``{Update on vector boson pair
  production at hadron colliders}'',} \textit{ Phys. Rev. D} \textbf{ 60}
  (1999) 113006,
  \href{http://dx.doi.org/10.1103/PhysRevD.60.113006}{\doi{10.1103/PhysRevD.60.113006}},
\href{http://www.arXiv.org/abs/hep-ph/9905386}{\texttt{arXiv:hep-ph/9905386}}.

\bibitem{Campbell:2010ff}
\hrefCMSnoop {}{J.~M. Campbell and R.~K. Ellis, ``{MCFM for the Tevatron and
  the LHC}'',} \textit{ Nucl. Phys. Proc. Suppl.} \textbf{ 205--206} (2010) 10,
  \href{http://dx.doi.org/10.1016/j.nuclphysbps.2010.08.011}{\doi{10.1016/j.nuclphysbps.2010.08.011}},
\href{http://www.arXiv.org/abs/1007.3492}{\texttt{arXiv:1007.3492}}.

\bibitem{Carli:2010rw}
T.~Carli\hrefCMSnoop {}{ {et~al.}, ``{A posteriori inclusion of parton density
  functions in NLO QCD final-state calculations at hadron colliders: the
  APPLGRID project}'',} \textit{ Eur. Phys. J. C} \textbf{ 66} (2010) 503,
  \href{http://dx.doi.org/10.1140/epjc/s10052-010-1255-0}{\doi{10.1140/epjc/s10052-010-1255-0}},
\href{http://www.arXiv.org/abs/0911.2985}{\texttt{arXiv:0911.2985}}.

\bibitem{Alekhin:2014irh}
\hrefCMSnoop {}{S.~Alekhin {et~al.}, ``{HERAFitter}'',} \textit{ Eur. Phys. J.
  C} \textbf{ 75} (2015) 304,
  \href{http://dx.doi.org/10.1140/epjc/s10052-015-3480-z}{\doi{10.1140/epjc/s10052-015-3480-z}},
\href{http://www.arXiv.org/abs/1410.4412}{\texttt{arXiv:1410.4412}}.

\bibitem{herafitter}
\hrefCMSnoop {}{}{HERAFitter} web site, \url{http://www.herafitter.org}.
  Accessed Aug 2015.

\bibitem{Gribov:1972ri}
\hrefCMSnoop {}{V.~N. Gribov and L.~N. Lipatov, ``{Deep inelastic $\Pe$-$\Pp$
  scattering in perturbation theory}'',} \textit{ Sov. J. Nucl. Phys.} \textbf{
  15} (1972)
438.

\bibitem{Altarelli:1977zs}
\hrefCMSnoop {}{G.~Altarelli and G.~Parisi, ``{Asymptotic freedom in parton
  language}'',} \textit{ Nucl. Phys. B} \textbf{ 126} (1977) 298,
\href{http://dx.doi.org/10.1016/0550-3213(77)90384-4}{\doi{10.1016/0550-3213(77)90384-4}}.

\bibitem{Curci:1980uw}
\hrefCMSnoop {}{G.~Curci, W.~Furmanski, and R.~Petronzio, ``{Evolution of
  parton densities beyond leading order: The non-singlet case}'',} \textit{
  Nucl. Phys. B} \textbf{ 175} (1980) 27,
\href{http://dx.doi.org/10.1016/0550-3213(80)90003-6}{\doi{10.1016/0550-3213(80)90003-6}}.

\bibitem{Furmanski:1980cm}
\hrefCMSnoop {}{W.~Furmanski and R.~Petronzio, ``{Singlet parton densities
  beyond leading order}'',} \textit{ Phys. Lett. B} \textbf{ 97} (1980) 437,
\href{http://dx.doi.org/10.1016/0370-2693(80)90636-X}{\doi{10.1016/0370-2693(80)90636-X}}.

\bibitem{Moch:2004pa}
\hrefCMSnoop {}{S.~Moch, J.~A.~M. Vermaseren, and A.~Vogt, ``{The three-loop
  splitting functions in QCD: the non-singlet case}'',} \textit{ Nucl. Phys. B}
  \textbf{ 688} (2004) 101,
  \href{http://dx.doi.org/10.1016/j.nuclphysb.2004.03.030}{\doi{10.1016/j.nuclphysb.2004.03.030}},
\href{http://www.arXiv.org/abs/hep-ph/0403192}{\texttt{arXiv:hep-ph/0403192}}.

\bibitem{Vogt:2004mw}
\hrefCMSnoop {}{A.~Vogt, S.~Moch, and J.~A.~M. Vermaseren, ``{The three-loop
  splitting functions in QCD: the singlet case}'',} \textit{ Nucl. Phys. B}
  \textbf{ 691} (2004) 129,
  \href{http://dx.doi.org/10.1016/j.nuclphysb.2004.04.024}{\doi{10.1016/j.nuclphysb.2004.04.024}},
\href{http://www.arXiv.org/abs/hep-ph/0404111}{\texttt{arXiv:hep-ph/0404111}}.

\bibitem{Botje:2010ay}
\hrefCMSnoop {}{M.~Botje, ``{QCDNUM: fast QCD evolution and convolution}'',}
  \textit{ Comput. Phys. Commun.} \textbf{ 182} (2011) 490,
  \href{http://dx.doi.org/10.1016/j.cpc.2010.10.020}{\doi{10.1016/j.cpc.2010.10.020}},
\href{http://www.arXiv.org/abs/1005.1481}{\texttt{arXiv:1005.1481}}.

\bibitem{Thorne:2006qt}
\hrefCMSnoop {}{R.~S. Thorne, ``{Variable-flavor number scheme for NNLO}'',}
  \textit{ Phys. Rev. D} \textbf{ 73} (2006) 054019,
  \href{http://dx.doi.org/10.1103/PhysRevD.73.054019}{\doi{10.1103/PhysRevD.73.054019}},
\href{http://www.arXiv.org/abs/hep-ph/0601245}{\texttt{arXiv:hep-ph/0601245}}.

\bibitem{Aaron:2009aa}
\hrefCMSnoop {}{{H1 and ZEUS Collaborations}, ``{Combined measurement and QCD
  analysis of the inclusive $\Pe^\pm \Pp$ scattering cross sections at
  HERA}'',} \textit{ JHEP} \textbf{ 01} (2010) 109,
  \href{http://dx.doi.org/10.1007/JHEP01(2010)109}{\doi{10.1007/JHEP01(2010)109}},
\href{http://www.arXiv.org/abs/0911.0884}{\texttt{arXiv:0911.0884}}.

\end{thebibliography}\endgroup

\cleardoublepage \appendix\section{The CMS Collaboration \label{app:collab}}\begin{sloppypar}\hyphenpenalty=5000\widowpenalty=500\clubpenalty=5000\textbf{Yerevan Physics Institute,  Yerevan,  Armenia}\\*[0pt]
V.~Khachatryan, A.M.~Sirunyan, A.~Tumasyan
\vskip\cmsinstskip
\textbf{Institut f\"{u}r Hochenergiephysik der OeAW,  Wien,  Austria}\\*[0pt]
W.~Adam, E.~Asilar, T.~Bergauer, J.~Brandstetter, E.~Brondolin, M.~Dragicevic, J.~Er\"{o}, M.~Flechl, M.~Friedl, R.~Fr\"{u}hwirth\cmsAuthorMark{1}, V.M.~Ghete, C.~Hartl, N.~H\"{o}rmann, J.~Hrubec, M.~Jeitler\cmsAuthorMark{1}, A.~K\"{o}nig, M.~Krammer\cmsAuthorMark{1}, I.~Kr\"{a}tschmer, D.~Liko, T.~Matsushita, I.~Mikulec, D.~Rabady, N.~Rad, B.~Rahbaran, H.~Rohringer, J.~Schieck\cmsAuthorMark{1}, R.~Sch\"{o}fbeck, J.~Strauss, W.~Treberer-Treberspurg, W.~Waltenberger, C.-E.~Wulz\cmsAuthorMark{1}
% ...
\vskip\cmsinstskip
% ...
1:~~Also at Vienna University of Technology, Vienna, Austria\\
% ...

\end{sloppypar}
\end{document}